\documentclass[journal]{IEEEtran}
\ifCLASSINFOpdf
\else
\fi
\usepackage{graphicx,cite,amssymb,amsmath,bm}
\usepackage{arydshln}
\usepackage{centernot}
\usepackage{float}
\usepackage{mathrsfs}
\usepackage{dsfont}
\usepackage{tikz}
\usepackage{xcolor}
\usepackage{etoolbox} 
\usepackage{booktabs} 
\usepackage[hyphens]{url} 
\usepackage{cite}
\tikzstyle{startstop} = [rectangle, rounded corners, minimum width=3cm, minimum height=1cm,text centered, draw=black, fill=red!30]
\tikzstyle{process} = [rectangle, minimum width=3cm, minimum height=1cm, text centered, draw=black, fill=blue!20]
\tikzstyle{decision} = [diamond, minimum width=3cm, minimum height=1cm, text centered, draw=black, fill=green!30]
\tikzstyle{arrow} = [thick,->,>=stealth]

 \usepackage{tikz}
 \usetikzlibrary{arrows.meta,positioning,calc}
\usepackage{subcaption}

\usepackage{comment,xcolor}

\newcounter{subassumption}[asu]

\makeatletter
\renewcommand{\p@subassumption}{\theasu}
\makeatother

\usepackage{color,xcolor}

\newtheorem{lemma}{\textbf{Lemma}}

\newtheorem{definition}{Definition}
\newtheorem{theorem}{\textbf{Theorem}}
\newtheorem{corollary}{Corollary}

\usepackage{mathtools}
\usepackage{epstopdf}

\usepackage[ruled,vlined,linesnumbered]{algorithm2e}

\begin{document}
	\title{PASS-Enabled Covert Communications With Distributed Cooperative Wardens}

\author{Ji He,~\IEEEmembership{Member,~IEEE,}
\thanks{J. He is with the School of Computer Science and Technology, Xidian University, Xi'an, 710071 China (e-mail: jihe@xidian.edu.cn).}
}

	\IEEEtitleabstractindextext{
		\begin{abstract}
This paper investigates PASS-enabled downlink covert communication in the presence of distributed surveillance, where multiple wardens perform signal detection and fuse their local binary decisions via majority-voting rule. We consider a dual-waveguide architecture that simultaneously delivers covert information and randomized jamming to hide the transmission footprint, incorporating three representative PASS power-radiation laws—general, proportional, and equal. To characterize the system-level detectability, we derive closed-form expressions for local false-alarm and miss-detection probabilities. By leveraging a probability-generating-function (PGF) and elementary-symmetric-polynomial (ESP) framework, combined with a breakpoint-based partition of the threshold domain, we obtain explicit closed-form characterizations of the system-level detection error probability (DEP) under non-i.i.d. majority-voting fusion. Building on this analytical framework, we formulate a robust optimization problem to maximize the average covert rate subject to covertness constraint. To solve the resulting nonconvex design, we develop an MM–BCD–SCA algorithm that produces tractable alternating updates for power/radiation variables and PA positions via convex surrogates and inner approximations of the DEP value function. Numerical results validate the theoretical analysis and demonstrate the impact of cooperative monitoring and PASS radiation laws on the covertness–rate tradeoff.
		\end{abstract}
		}
	
	\maketitle
	\IEEEdisplaynontitleabstractindextext
	\IEEEpeerreviewmaketitle
	
\section{INTRODUCTION}\label{sec:Introduction}

6G is expected to move beyond peak-rate gains and deliver uniform quality of service (QoS)—high reliability, low latency, and scalable connectivity—across heterogeneous and dynamic environments. As envisioned 6G use cases proliferate, a persistent practical bottleneck shifts from spectrum scarcity and baseband processing to the reliable delivery of electromagnetic energy to desired locations despite mobility, blockage, and deployment constraints \cite{wang2023road,shen2023five}. This motivates reconfigurable antenna architectures that can reshape large-scale propagation with practical deployments, among which pinching-antenna systems (PASS) have been proposed as a ``near-wire" architecture \cite{liu2025pinching}. In PASS, a dielectric waveguide acts as the guided transmission backbone, while small dielectric elements, termed pinching antennas (PAs), are placed at selected locations to locally couple guided waves to free-space radiation, thereby creating reconfigurable transmit/receive apertures along the waveguide \cite{suzuki2022pinching}.

From a communications-performance perspective, PASS alters the link budget and channel geometry by guiding most of the energy along the waveguide and radiating it only at deliberately selected locations. By activating PAs near the intended users, PASS can reinforce dominant propagation paths and mitigate large-scale attenuation, while the ability to reconfigure the active radiation points provides a practical mechanism to bypass unfavorable propagation conditions without physically redesigning a fixed antenna array \cite{liu2025pinching1}. Moreover, simultaneous activation of multiple PAs provides an additional spatial degree of freedom, often termed pinching beamforming, which complements conventional precoding by shaping both small-scale phases and large-scale channel gains through PA placement \cite{zhang2025two}. Recent designs that jointly optimize transmit beamforming and pinching beamforming (e.g., via two-timescale formulations) report notable sum-rate gains over baselines with fixed or non-optimized pinching configurations. Analytical results further indicate that LoS blockage can suppress co-channel interference and widen the performance gap relative to conventional antennas in interference-limited multiuser regimes \cite{ding2025blockage}.

Beyond throughput and coverage, PASS is particularly attractive for physical-layer security (PLS) because it offers geometry-driven control variables that directly shape the channel advantage required for information-theoretic secrecy. Classical PLS establishes that secrecy is achievable when the legitimate link is made sufficiently stronger than the eavesdropper’s link, motivating beamforming, artificial noise (AN), and user-centric channel-shaping strategies in multiuser wireless systems. PASS enriches this toolbox via location-selective radiation: because energy is primarily guided and radiated only at chosen PA locations, the transmitter can promote constructive superposition at the legitimate user, reduce unnecessary exposure, and induce unfavorable combining at an eavesdropper via PA placement. Recent PASS-enabled secure-communication frameworks operationalize this intuition by optimizing PA locations and—under multi-waveguide architectures—jointly designing pinching beamforming, artificial noise, and power allocation to maximize the secrecy rate subject to practical architectural constraints \cite{xu2025generalized}.

Early studies on PASS-enabled PLS primarily establish how pinching beamforming creates secrecy gains by introducing a geometry-driven degree of freedom beyond conventional baseband precoding. In \cite{sun2025physical}, the authors formulate secrecy beamforming for single-user and multiuser PASS, derive closed-form baseband beamformers, and optimize PA activation locations via gradient-based and FP–BCD methods to maximize the weighted secrecy sum rate. Focusing on a minimal single-PA wiretap link over a dielectric waveguide, the work \cite{badarneh2025} provides analytical expressions for the average secrecy capacity, secrecy outage probability, and related metrics, thereby clarifying the impact of PA placement. Building on these foundations, the works \cite{papanikolaou2025} and \cite{zhu2025pinching} incorporate artificial noise (AN) to enhance security in single- and multi-waveguide PASS, respectively. Beyond these canonical downlink wiretap formulations, more recent studies explore PASS-specific security mechanisms and service modes. In \cite{wang2025pinching}, discretely pre-installed PAs are assumed, and secrecy-rate maximization is demonstrated via waveguide-enabled amplitude and phase control. Dual-waveguide PASS is examined in \cite{lu2025dual}, which compares parallel and orthogonal placement strategies and proposes a two-stage algorithm (PSO followed by SCA) to optimize both the secure sum rate and secure energy efficiency. For group-oriented services, the work \cite{shan2025secure} studies secure multicast in single- and multi-group PASS and develops alternating/MM-based designs to maximize the secrecy multicast rate. At the waveform/modulation layer, the authors combine PASS with index and directional modulation to further harden the signal structure against interception in \cite{zhong2025physical}. To mitigate internal eavesdropping induced by SIC in PASS-aided NOMA system,  the work \cite{chi2025pinching} proposes adaptive PA power and coupling-length control strategies.

While early PASS studies mainly focus on secrecy-oriented PLS, a closely related yet stricter requirement in adversarial settings is physical-layer covert communication, which aims to make the transmission undetectable to a warden while maintaining a nontrivial communication rate. In this direction, \cite{jiang2025pinching} first investigates how PA location programmability can be leveraged to improve covert throughput under a covertness constraint. To move beyond static-warden assumptions, \cite{jiang2025} develops a joint design of beamforming, AN, and PA positioning to handle time-varying adversary states. PASS-enabled covertness is further extended to backscatter communications in \cite{wang2025uplink}, where numerical results suggest that PASS can mitigate the ``double near–far" effect while maintaining low detectability against randomly located adversaries.

Recent studies on PASS-enabled covert communication show that PA location programmability can be exploited to enhance covertness. However, practical surveillance is often distributed: multiple wardens independently sense the spectrum and subsequently fuse their local decisions to reach a global verdict \cite{eryigit2013channel,akyildiz2011,fernando2019}. In this setting, ensuring covertness requires controlling the \emph{system-level} detection error probability (DEP) under cooperative fusion, rather than relying on a single-warden metric. For PASS networks, this task is further complicated by two intrinsic features: (i) the warden-side detector statistics are generally \emph{non-identically distributed} due to heterogeneous warden locations/channels, rendering the resulting fusion DEP analytically challenging; and (ii) PASS exhibits hardware-coupled power–radiation behavior along the waveguide, so the radiated power observed at each warden depends jointly on the PA configuration and the waveguide coupling mechanism. Our key contributions are summarized below.

\begin{itemize}
  \item We study a dual-waveguide PASS architecture in which both the communication and jamming waveguides are equipped with PAs. Each waveguide can be partitioned via waveguide division (WD) to enable location-selective radiation/collection and structured PA deployment. We further incorporate three representative PASS power--radiation laws: the general, proportional, and equal models.

  \item Under the above three power-radiation models, we derive closed-form expressions of each warden's false-alarm and miss-detection probabilities as functions of the detection threshold. By leveraging a PGF/ESP-based representation, we further obtain explicit closed-form system DEP expressions for majority-voting fusion in the non-i.i.d. case, enabled by a systematic partition of the threshold domain through warden-dependent breakpoints.

  \item Building on the developed analytical framework, we formulate a robust average covert-rate maximization problem subject to a worst-case covertness constraint and practical power/placement constraints. We develop an MM--BCD--SCA algorithm that yields tractable alternating subproblems for the joint optimization of power/radiation variables and PA locations, and we use the resulting solutions to quantify how cooperative monitoring and PASS radiation laws reshape the covertness--rate tradeoff.

  \item Extensive numerical results demonstrate how the cooperative-warden DEP structure and the considered radiation models affect the covertness--rate tradeoff. The results also provide actionable guidelines for configuring WD and selecting radiation models under distributed spectrum monitoring.
\end{itemize}

The remainder of this paper is organized as follows. Section~\ref{sec:sys_model} presents the system model. Section~\ref{sec:performance} defines and analyzes the performance metric. Section~\ref{sec:system_DEP} derives a closed-form expression for the system-level DEP and addresses the associated performance optimization. Numerical results are provided in Section~\ref{sec:numer_results}, and Section~\ref{sec:conclusion} concludes the paper.


\section{SYSTEM MODEL AND PRELIMINARIES}
\label{sec:sys_model}

In this section, we first describe the overall system architecture and channel model, then detail the power radiation models for the pinching antennas and the warden detection scheme that will be used in the subsequent performance analysis.

\subsection{System Model}

\begin{figure}[t]
\centering
\includegraphics[width=1\linewidth]{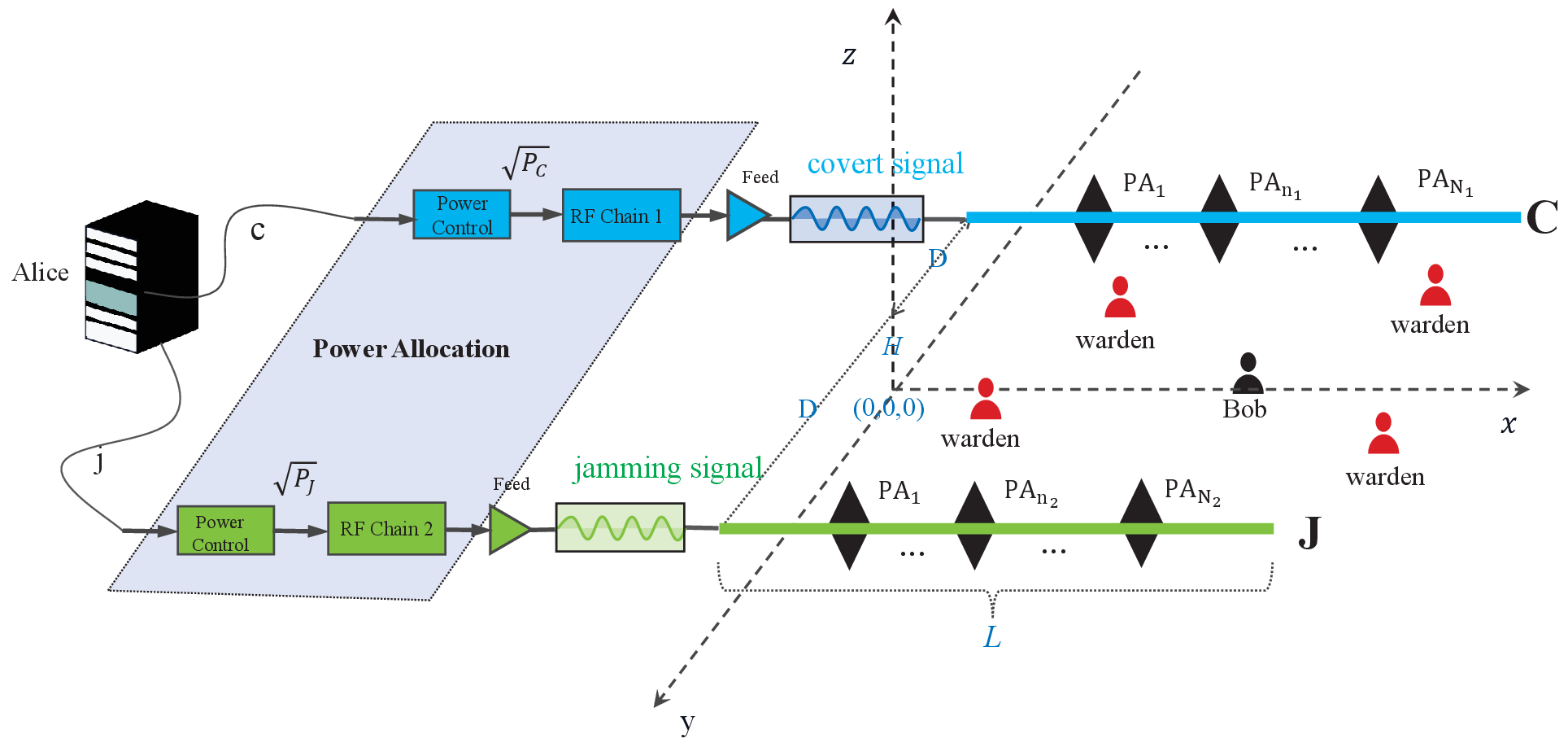}
\caption{Illustration of a PASS-enabled covert communication system with dual waveguides.}
\label{fig:system_model}
\end{figure}

As illustrated in Fig.~\ref{fig:system_model}, we consider a downlink covert communication system where a base station (BS), referred to as Alice, intends to communicate covertly with a legitimate user Bob in the presence of $M$ wardens, denoted by $w_1,w_2,\ldots,w_M$. Alice is equipped with two parallel dielectric waveguides of length $L$ that share a total of $N$ PAs. By employing the WD technique \cite{zhu2025pinching}, each waveguide is connected to a single RF chain and carries one data stream, so that the two waveguides can simultaneously deliver covert information and artificial jamming. The first waveguide, denoted as waveguide $C$, is dedicated to transmitting covert information-bearing signals and contains $N_1$ PAs. The second waveguide, denoted as waveguide $J$, generates intentional jamming signals to confuse the wardens and contains the remaining $N_2$ PAs, with $N_1+N_2=N$. Let $P_{\max}$ denote Alice's maximum transmit power. The powers allocated to covert communication and jamming are denoted by $P_C$ and $P_J$, respectively, and satisfy $P_C+P_J\leq P_{\max}$. To enhance covertness, the instantaneous jamming power $P_J$ is modeled as a random variable uniformly distributed over $[0,P_J^{\max}]$, i.e.,
\begin{equation}\label{eq:pdf_j}
 f_{P_J}(x) =
\begin{cases}
\dfrac{1}{P_J^{\max}}, & 0 \le x \le P_J^{\max}, \\
0, & \text{otherwise}.
\end{cases}
\end{equation}
which randomizes the interference observed at the wardens.

We establish a three-dimensional Cartesian coordinate system in which the two waveguides are placed parallel to the $x$-axis at a height $H$ above the ground, and their perpendicular distance to the $x$-axis is $D$. Accordingly, the coordinates of the PAs on waveguide~$C$ are $[x^{C}_{n_1},-D,H]^T$ for $n_1\in{1,\ldots,N_1}$, and those on waveguide~$J$ are $[x^{J}_{n_2},D,H]^T$ for $n_2\in{1,\ldots,N_2}$. Bob and the wardens are randomly located on the ground plane. The positions of Bob and warden $m$ are denoted by $[x_b,y_b,0]^T$ and $[x_w,y_w,0]^T$, respectively.

\subsection{Channel Modeling}

In this subsection, we characterize the equivalent channels between Alice and each node in the network. Let $\boldsymbol{x}_C=[x^C_1,x^C_2,\ldots,x^C_{N_1}]^T$ denote the vector collecting the $x$-coordinates of all PAs on waveguide~$C$, and let $\boldsymbol{x}_J=[x^J_1,x^J_2,\ldots,x^J_{N_2}]^T$ denote the corresponding vector for waveguide~$J$. The phase shifts caused by propagation along the dielectric waveguides are represented by
\begin{equation}
\boldsymbol{\omega}_C = e^{-\mathrm{j}\frac{2\pi}{\lambda_g}\boldsymbol{x}_C}, \quad
\boldsymbol{\omega}_J = e^{-\mathrm{j}\frac{2\pi}{\lambda_g}\boldsymbol{x}_J},
\end{equation}
where $\frac{2\pi}{\lambda_g}$ is the effective propagation constant in the waveguide, $\lambda_g = \lambda / n_{\text{eff}}$, $\lambda$ is the carrier wavelength in free space, and $n_{\text{eff}}$ is the effective refractive index of the dielectric waveguide \cite{ding2025flexible}.

Let $k\in\{\text{Bob},w_1,w_2,\ldots,w_M\}$ index Bob and the wardens. Under line-of-sight propagation, the free-space channel vectors from waveguides $C$ and $J$ to user $k$ are given by
\begin{align}
\bm{\mathrm{h}}_k(\boldsymbol{x}_C)
&= \left[ \frac{\eta^{1/2} e^{-\mathrm{j}\frac{2\pi}{\lambda} \| d^C_{k,1} \|}}{\|d^C_{k,1} \|}, \ldots,
\frac{\eta^{1/2} e^{-\mathrm{j}\frac{2\pi}{\lambda} \| d^C_{k,N_1} \|}}{\| d^C_{k,N_1} \|} \right]^T, \\
\bm{\mathrm{h}}_k(\boldsymbol{x}_J)
&= \left[ \frac{\eta^{1/2} e^{-\mathrm{j}\frac{2\pi}{\lambda} \| d^J_{k,1} \|}}{\| d^J_{k,1} \|}, \ldots,
\frac{\eta^{1/2} e^{-\mathrm{j}\frac{2\pi}{\lambda} \| d^J_{k,N_2} \|}}{\| d^J_{k,N_2} \|} \right]^T,
\end{align}
where $\eta = \frac{\lambda^2}{16\pi^2}$ represents the free-space propagation constant and $\|d^{C}_{k,n_1}\|$ (resp. $\|d^{J}_{k,n_2}\|$) denotes the Euclidean distance between user $k$ at $(x_k,y_k,0)$ and the $n_1$-th PA on waveguide~$C$ (resp. the $n_2$-th PA on waveguide~$J$). These vectors capture the path-loss and phase shift from each PA to user $k$.

By combining the effects of waveguide and free-space propagation, the equivalent complex baseband channel coefficient between Alice and user $k$ can be modeled as \begin{equation}\label{eq:channel gain}
h_k(\bm{x}_C)=\bm{\mathrm{h}}_k^T(\bm{x}_C)\boldsymbol{\omega}_C,\quad h_k(\boldsymbol{x}_J)=\bm{\mathrm{h}}_k^T(\boldsymbol{x}_J)\boldsymbol{\omega}_J.
\end{equation}

\subsection{Power Radiation Models}
	
The power radiated by each PA is affected by the power exchange (coupling) at all preceding PAs along the same waveguide. In this work, we consider three typical power radiation models \cite{liu2025pinching}: the general power radiation model, the proportional power radiation model, and the equal power radiation model.
	
\textbf{General Power Radiation Model:}
In the general model, the radiated power of each PA depends on the coupling coefficients of all upstream PAs. The power radiated by the $n_1$-th PA on waveguide~$C$ and the $n_2$-th PA on waveguide~$J$ is given by
	\begin{align}
		 P_{C,n_1}^G &= \rho_{C,n_1}^G P_C \,\, \text{where}\,\, \rho_{C,n_1}^G = \delta_{C,n_1}^2 \prod_{i=1}^{n_1 - 1} (1 - \delta_{C,i}^2),   \\
		 P_{J,n_2}^G &= \rho_{J,n_2}^G P_J \,\, \text{where}\,\,\rho_{J,n_2}^G = \delta_{J,n_2}^2 \prod_{i=1}^{n_2 - 1} (1 - \delta_{J,i}^2),
	\end{align}
	where $\delta_{C,n_1} = \sin(\kappa_{C,n_1} L_{C,n_1})$ and $\delta_{J,n_2} = \sin(\kappa_{J,n_2} L_{J,n_2})$ represent the coupling strength determined by the coupling coefficient $\kappa$ and the coupling length $L$.
	
\textbf{Proportional Power Radiation Model:}
In the proportional model, each PA radiates a fixed fraction of the remaining power after all preceding PAs have radiated, so that the power is allocated proportionally rather than uniformly. The power radiated by the $n_1$-th PA on waveguide~$C$ and the $n_2$-th PA on waveguide~$J$ is expressed as
	\begin{align}
		 P_{C,n_1}^P &= \rho_{C,n_1}^P P_C\,\, \text{where}\,\, \rho_{C,n_1}^P = \delta_C^2 (1 - \delta_C^2)^{n_1 - 1},  \\
		 P_{J,n_2}^P& = \rho_{J,n_2}^P P_J\,\, \text{where}\,\, \rho_{J,n_2}^P = \delta_J^2 (1 - \delta_J^2)^{n_2 - 1},
	\end{align}
where $\delta_C^2$ and $\delta_J^2$ denote the fractions of power that each PA extracts from waveguides~$C$ and~$J$, respectively, and satisfy $\delta_C^2,\delta_J^2 \in (0,1)$.
	
	\textbf{Equal Power Radiation Model}: In the equal model, the power on each waveguide is evenly allocated among its PAs, so that every PA radiates the same fraction of the total power. The power radiated by the $n_1$-th PA on waveguide~$C$ and the $n_2$-th PA on waveguide~$J$ is given by
	\begin{align}
	\quad P_{C,n_1}^E &\!=\! \rho_{C,n_1}^E P_C \,\, \text{where}\,\,\rho_{C,n_1}^E \!\triangleq \!\rho_{C},\,  0 < \rho_{C} \leq \frac{1}{N_1}, \\
		  P_{J,n_2}^E &= \rho_{J,n_2}^E P_J \,\, \text{where}\,\, \rho_{J,n_2}^E \triangleq \rho_{J}, \, 0 < \rho_{J} \leq \frac{1}{N_2}.
	\end{align}
where $\rho_C$ and $\rho_J$ are the constant power fractions allocated to each PA on waveguides~$C$ and~$J$, respectively.

\subsection{Warden Detection Scheme}

In the considered system, multiple wardens independently perform spectrum sensing and form local binary decisions, which are then combined using a majority-voting fusion rule (i.e., the minority follows the majority) to determine whether covert communication is present.
Since the local decisions are conditionally independent given each hypothesis, we first characterize the underlying binary hypothesis test at an individual warden and analyze its local decision rule, which will later be used to evaluate the global detection performance under majority voting.

Under the null hypothesis $H_0$, Alice does not transmit covert information and only the jamming signal is present, whereas under the alternative hypothesis $H_1$ both the covert signal and the jamming signal are transmitted.
The received signal at warden $w_m$ in the $i$-th symbol interval can be expressed as
\begin{equation}\label{eq:signal_warden}
y_{w_m}\!(i) \!=\!
\begin{cases}
\bm{\mathrm{h}}_{w_m}^T(\boldsymbol{x}_J)\sqrt{P_{J}}\mathbf{\Theta}^*_J \boldsymbol{\omega}_J s_j(i)
+ n_{w}(i), & H_0 \\
\bm{\mathrm{h}}_{w_m}^T(\boldsymbol{x}_C)\sqrt{P_{C}}\mathbf{\Theta}^*_C\boldsymbol{\omega}_C s_c(i) \\
 \,+ \bm{\mathrm{h}}_{w_m}^T(\!\boldsymbol{x}_J)\sqrt{P_{J}}\mathbf{\Theta}^*_J\boldsymbol{\omega}_J s_j(i)
 + n_{w}(i), & H_1
\end{cases}
\end{equation}
where $s_j(i)$ and $s_c(i)$ denote the jamming and covert signals, respectively, and $n_{w_m}(i)\sim\mathcal{CN}(0,\sigma_w^2)$ is the additive white Gaussian noise (AWGN) at warden $w_m$. Here, $\mathbf{\Theta}_C^* = \operatorname{diag}\bigl(\sqrt{\rho_{C,1}^*},\ldots,\sqrt{\rho_{C,N_1}^*}\bigr)$ and $\mathbf{\Theta}_J^* = \operatorname{diag}\bigl(\sqrt{\rho_{J,1}^*},\ldots,\sqrt{\rho_{J,N_2}^*}\bigr)$ are the diagonal matrices describing the per-PA power radiation along waveguides $C$ and $J$, and $*\in \{G,P,E\}$ specifies the adopted power radiation model.

According to the Neyman--Fisher factorization theorem, the accumulated received energy $\sum_{i=1}^n |y_{w_m}(i)|^2$ is a sufficient statistic for detecting the presence of the covert signal at warden $w_m$ \cite{sobers2017covert}. Therefore, the optimal energy detector at warden $w_m$ can be written as
\begin{equation} \label{eq:BHT}
Y_{w_m}=\frac{1}{n} \sum_{i=1}^n |y_{w_m}(i)|^2 \underset{\mathcal{D}_0}{\overset{\mathcal{D}_1}{\gtrless}} \tau,
\end{equation}
where $n$ denotes the number of complex samples collected in one detection interval, $\tau>0$ is the decision threshold, and $\mathcal{D}_0$ and $\mathcal{D}_1$ denote the local decisions corresponding to hypotheses $H_0$ (no covert transmission) and $H_1$ (covert transmission), respectively.


\section{Performance Analysis}
\label{sec:performance}
In this section, we first derive the instantaneous covert transmission rate, then characterize the local detection behavior of each warden, obtaining the system-level DEP under the majority-voting fusion rule. Finally, we investigate the performance optimal problem.

\subsection{Covert Transmission Rate}

In this subsection, we characterize the transmission performance at Bob when Alice performs covert communication in the presence of artificial jamming. The received signal at Bob in the $i$-th symbol interval can be written as
\begin{align}\label{eq:signal_bob}
y_{b}(i)= &\bm{\mathrm{h}}_{b}^T(\boldsymbol{x}_C)\sqrt{P_{C}}\mathbf{\Theta}_C^* \boldsymbol{\omega}_{C} s_c(i) \nonumber \\
&+ \bm{\mathrm{h}}_{b}^T(\boldsymbol{x}_J)\sqrt{P_{J}}\mathbf{\Theta}_J^* \boldsymbol{\omega}_J s_j(i)
+ n_b(i),
\end{align}
where $\mathbb{E}\{|s_c(i)|^2\}=\mathbb{E}\{|s_j(i)|^2\}=1$, and $n_b(i)\sim\mathcal{CN}(0,\sigma_b^2)$ is the AWGN at Bob.

So, the instantaneous covert transmission rate at Bob is $R_c=\log_2(1+\gamma_b)$ where $\gamma_b$ is the SINR at Bob determined as
	\begin{align} \label{eq:SINR}
		\gamma_b &= \frac{\left|\bm{\mathrm{h}}_{b}^T(\boldsymbol{x}_{C})\sqrt{P_{C}}\mathbf{\Theta}^*_C\boldsymbol{\omega}_{C}\right|^2}{\left|\bm{\mathrm{h}}_{b}^T(\boldsymbol{x}_{J})\sqrt{P_{J}}\mathbf{\Theta}^*_J\boldsymbol{\omega}_{J} s_j(i)\right|^2+\delta^2_b} \nonumber \\
&=\frac{\left| \sum_{n_1=1}^{N_1} \!\sqrt{\!P_{C,n_1}^*} e^{-\mathrm{j}\frac{2\pi}{\lambda_g} x_{C,n_1}}  \frac{\eta^{1/2} e^{-\mathrm{j}\frac{2\pi}{\lambda} \| \boldsymbol{d}^c_{k,n_1}\! \|}}{\| \boldsymbol{d}^c_{k,n_1} \|} \right|^2}{\left| \sum_{n_2=1}^{N_2}\! \sqrt{\!P_{J,n_2}^*} e^{-\mathrm{j}\frac{2\pi}{\lambda_g} x_{J,n_2}} \frac{\eta^{1/2} e^{-\mathrm{j}\frac{2\pi}{\lambda} \| \boldsymbol{d^J}_{k,n_2} \|}}{\| \boldsymbol{d}^J_{k,n_2} \|} \right|^2 \!\!\!+ \sigma_b^2}
	\end{align}
where $P_{C,n_1}^*$ and $P_{J,n_2}^*$ denote the radiated powers of the $n_1$-th and $n_2$-th PAs on waveguides $C$ and $J$, respectively, under the adopted power-radiation model $*\in\{G,P,E\}$.

\subsection{Local Detection Performance}

Within the framework of binary hypothesis testing, a miss detection at warden $w_m$ occurs when $w_m$ decides $\mathcal{D}_0$ under $H_1$, whereas a false alarm occurs when it decides $\mathcal{D}_1$ under $H_0$. Accordingly, the local false alarm and miss detection probabilities are defined as $P_{\mathrm{fa},m} = \mathrm{P_r}\{ \mathcal{D}_1 | H_0 \}, P_{\mathrm{md},m} = \mathrm{P_r}\{\mathcal{D}_0 | H_1 \}$. Based on the energy detector in \eqref{eq:BHT}, we can derive closed-form expressions for these probabilities as functions of the decision threshold~$\tau$.

\begin{lemma}\label{eq:lemm_local-detec}
In the considered PASS-enabled covert communication system, the false alarm and miss detection probabilities at warden $w_m$ for an arbitrary detection threshold $\tau$ are given by
\begin{align}
P_{\text{fa},m} &=
\begin{cases} \label{eq:fa}
1, & \tau < \sigma_{w}^2, \\
\dfrac{\alpha_{1,m} - \tau}{P_J^{\max} A_{J,m}}, & \sigma_{w}^2 \leq \tau \leq \alpha_{1,m}, \\
0, & \text{otherwise},
\end{cases} \\
P_{\text{md},m} &=
\begin{cases} \label{eq:md}
0, & \tau < \alpha_{2,m}, \\
\dfrac{\tau - \alpha_{2,m}}{P_J^{\max} A_{J,m}}, & \alpha_{2,m} \leq \tau \leq \alpha_{3,m}, \\
1, & \text{otherwise},
\end{cases}
\end{align}
where $\alpha_{1,m} = \sigma_{w}^2 + P_J^{\max} A_{J,m}$, $\alpha_{2,m} = \sigma_{w}^2 + P_C A_{C,m}$, $\alpha_{3,m} = \alpha_{1,m} + \alpha_{2,m} - \sigma_{w}^2$. Here, $A_{J,m} \triangleq \Bigl|\bm{\mathrm{h}}_{w_m}^T(\boldsymbol{x}_{J}) \mathbf{\Theta}_J^* \boldsymbol{\omega}_{J}\Bigr|^2$ and $A_{C,m} \triangleq \Bigl|\bm{\mathrm{h}}_{w_m}^T(\boldsymbol{x}_{C}) \mathbf{\Theta}_C^* \boldsymbol{\omega}_{C}\Bigr|^2$ denote the effective channel gains from the jamming and covert waveguides to warden $w_m$, respectively.
\end{lemma}

\begin{IEEEproof}
Substituting \eqref{eq:signal_warden} into \eqref{eq:BHT}, the test statistic at warden $w_m$ under $H_0$ can be written as
\begin{align}
Y_{w_m} | H_0
&= \frac{1}{n} \sum_{i=1}^n \bigl| \bm{\mathrm{h}}_{w_m}^T(\boldsymbol{x}_{J}) \sqrt{P_{J}} \mathbf{\Theta}_J^* \boldsymbol{\omega}_{J} s_j(i) + n_w(i) \bigr|^2 \nonumber \\
&= P_J A_{J,m} + \sigma_w^2,
\end{align}
where we have used the unit-power assumption on $s_j(i)$ and the independence between signal and noise. Hence,
\begin{align} \label{eq:pfa_cal}
P_{\text{fa},m}
&= \Pr\{ Y_{w_m} > \tau \mid H_0 \}= \Pr\left\{ P_J A_{J,m} + \sigma_w^2 > \tau \right\} \nonumber \\
&= \Pr\left\{ P_J > \frac{\tau - \sigma_{w}^2}{A_{J,m}} \right\}.
\end{align}
Using the PDF of $P_J$ in \eqref{eq:pdf_j} and performing straightforward integration yields \eqref{eq:fa}.

Similarly, under $H_1$ we obtain
\begin{align}
Y_{w_m} | H_1
&= \frac{1}{n} \sum_{i=1}^n \bigl| \bm{\mathrm{h}}_{w_m}^T(\boldsymbol{x}_{C}) \sqrt{P_{C}} \mathbf{\Theta}_C^* \boldsymbol{\omega}_{C} s_c(i) \nonumber \\
&\quad\quad\quad\ + \bm{\mathrm{h}}_{w_m}^T(\boldsymbol{x}_{J}) \sqrt{P_{J}} \mathbf{\Theta}_J^* \boldsymbol{\omega}_{J} s_j(i) + n_w(i) \bigr|^2 \nonumber \\
&= P_C A_{C,m} + P_J A_{J,m} + \sigma_w^2,
\end{align}
which leads to
\begin{align}\label{eq:pmd_cal}
P_{\text{md},m}
&\!=\! \Pr\{ Y_{w_m} \!\!<\! \tau | H_1 \}\!=\! \Pr\left\{\! P_C A_{C,m} \!+ \!P_J A_{J,m} \!+ \!\sigma_w^2 \!< \!\tau \right\} \nonumber \\
&\!= \Pr\left\{ P_J < \frac{\tau - \sigma_{w}^2 - P_C A_{C,m}}{A_{J,m}} \right\}.
\end{align}
Again, substituting the PDF of $P_J$ in \eqref{eq:pdf_j} into \eqref{eq:pmd_cal} and integrating over the support of $P_J$ yields \eqref{eq:md}, which completes the proof.
\end{IEEEproof}

Although Lemma~\ref{eq:lemm_local-detec} completely characterizes the local detection performance of each warden, the system-level DEP under the majority-voting fusion rule is more involved. This is because the local statistics $\{Y_{w_m}\}$ are independent but in general not identically distributed, as the parameters $(\alpha_{1,m},\alpha_{2,m},\alpha_{3,m})$ depend on the individual channels and locations of the wardens. As a result, the overall detection performance cannot be analyzed by simply invoking binomial models for identically distributed detectors. In the following, we introduce several auxiliary definitions and theorems to facilitate the analysis of the system-level DEP.

\subsection{System-Level DEP Analysis}

\begin{definition}[Probability generating function]
Let $S$ be a nonnegative, integer-valued random variable. The probability
generating function (PGF) of $S$ is defined as
\begin{equation}
G_S(z) \triangleq \mathbb{E}\!\left[z^{S}\right]
= \sum_{i=0}^{\infty} \Pr(S=i)\,z^i,
\qquad |z|\le 1.
\label{eq:pgf_def}
\end{equation}
\end{definition}

The PGF compactly encodes all probabilities $\Pr(S=i)$ in the coefficients of the power series in~\eqref{eq:pgf_def}. The following property links
the tail probability of $S$ to the coefficients of its PGF.

\begin{lemma}
\label{lem:tail_coeff}
Let $S$ be as in Definition~\ref{eq:pgf_def} with PGF $G_S(z)$. Then, for any integer threshold $T \ge 0$,
\begin{equation}
\Pr(S \ge T) = \sum_{i=T}^{\infty} [z^i] G_S(z),
\label{eq:tail_coeff}
\end{equation}
where $[z^i] G_S(z)$ denotes the coefficient of $z^i$ in the series expansion of $G_S(z)$.
\end{lemma}

\begin{IEEEproof}
By definition, $G_S(z) = \sum_{i=0}^{\infty} \Pr(S=i)\,z^i$. The operator $[z^i]$ simply extracts the coefficient in front of $z^i$, i.e.,
$[z^i]G_S(z)=\Pr(S=i)$. Therefore, $\Pr(S \ge T)= \sum_{i=T}^{\infty}\Pr(S=i)= \sum_{i=T}^{\infty}[z^i]G_S(z)$.
\end{IEEEproof}

Building on Lemma~\ref{lem:tail_coeff}, since each warden independently determines whether covert communication is present, the overall DEP under the majority-voting fusion rule can be characterized through the PGFs of the individual false-alarm and miss-detection probabilities,  $P_{\mathrm{fa},m}$ and $P_{\mathrm{md},m}$, respectively, as shown below.

\begin{theorem}
\label{thm:majority_dep_pgf}
 Consider a PASS-enhanced covert communication system with $M$ wardens using the majority voting fusion rule, the DEP $P_{dep}$ can be determined by
\begin{equation}
P_{dep}
=\sum_{i=T}^{M}[z^i]\,G_{\mathrm{FA}}(z)
+\sum_{i=0}^{T-1}[z^i]\,G_{\mathrm{D}}(z).
\label{eq:dep_majority_pgf}
\end{equation}
where $T=\lfloor M/2\rfloor+1$, and
\begin{align}
G_{\mathrm{FA}}(z)&=\prod_{m=1}^M [(1-P_{\mathrm{fa},m})+P_{\mathrm{fa},m}z] \label{eq:e_fa}
=\sum_{i=0}^{M} e_i^{(\mathrm{FA})}z^i,\\
G_{\mathrm{D}}(z)&=\!\prod_{m=1}^M [P_{\mathrm{md},m}+(1-P_{\mathrm{md},m})z] \label{eq:e_d}
=\!\sum_{i=0}^{M} e_i^{(\mathrm{D})}z^i.
\end{align}
\end{theorem}

\begin{IEEEproof}
Let $S_X=\sum_{m=1}^M X_m$ and $S_Y=\sum_{m=1}^M Y_m$ denote the number of local alarms under $H_0$ and the number of correct
detections under $H_1$, respectively. Applying Lemma~\ref{lem:tail_coeff} to the sets $\{X_m\}$ with $p_m=P_{\mathrm{fa},m}$ and
$\{Y_m\}$ with $p_m=1-P_{\mathrm{md},m}$, we can obtain the false alarm probability of the considered system as $P_{\mathrm{FA}}
=\Pr(S_X\ge T) =\sum_{i=T}^{M}[z^i]G_{\mathrm{FA}}(z)$. Note that the system missed detection event under majority voting is the number of correct detections is less than half. Thus, the miss detection probability is given as $P_{\mathrm{MD}}
=\Pr(S_Y\le T-1)=\sum_{i=0}^{T-1}[z^i]G_{\mathrm{D}}(z)$.
Summing these two equations immediately gives \eqref{eq:dep_majority_pgf}.
\end{IEEEproof}

Theorem~\ref{thm:majority_dep_pgf} provides a compact representation of the overall DEP $P_{\mathrm{dep}}$ through the PGFs $G_{\mathrm{FA}}(z)$ and $G_{\mathrm{D}}(z)$. However, to gain deeper analytical insight and facilitate subsequent performance evaluation, it is necessary to explicitly express the coefficients of these generating functions, i.e., $e_i^{(\mathrm{FA})}z^i$ and $e_i^{(\mathrm{D})}$. These coefficients correspond to the probabilities that exactly $i$ wardens report a false alarm or correctly detect the transmission, respectively. To derive these coefficients, consider the product term $\prod_{m=1}^M \left[(1 - P_{\mathrm{fa},m}) + P_{\mathrm{fa},m} z\right]$ in (\ref{eq:e_fa}), Each factor inside the parentheses represents the two possible outcomes of the $m$-th warden—either no false alarm $1 - P_{\mathrm{fa},m}$ or a false alarm $P_{\mathrm{fa},m} z$. When this product is expanded, every term in the resulting polynomial corresponds to a unique combination of false-alarm outcomes across all wardens.

\begin{definition} The $k$th elementary symmetric polynomial (ESP) of a vector $\mathbf{x}=(x_1,\ldots,x_M)$ be defined as
\begin{equation}
\xi_k(\mathbf{x}) \triangleq \sum_{\substack{S\subseteq\{1,\ldots,M\},|S|=k}}
\prod_{m\in S} x_m, \, k=\{0,1,\ldots,M\},
\label{eq:esp_def}
\end{equation}
with the convention $\xi_0(\mathbf{x})\equiv 1$.
\end{definition}

To systematically enumerate all such combinations, we apply EGF in \eqref{eq:e_fa} and \eqref{eq:e_d}, we can obtain the following corollary.

\begin{corollary} For all $i=0,1,\ldots,M$, the PGF coefficients admit the ESP-based representations
\begin{align}
e_i^{(\mathrm{FA})}
&\!=\!
\sum_{k=i}^{M}
(-1)^{k+i}
\binom{k}{i}\,
\xi_k\!\big(P_{\mathrm{fa},1},\ldots,P_{\mathrm{fa},M}\big),
\label{eq:ei_fa_esp}\\[1mm]
e_i^{(\mathrm{D})}
&\!=\!
\sum_{k=i}^{M}
(-1)^{k+i}
\binom{k}{i}\,
\xi_k\!\big(1\!-\!P_{\mathrm{md},1},\ldots,1\!-\!P_{\mathrm{md},M}\big).
\label{eq:ei_d_esp}
\end{align}
\label{coro:ei}
\end{corollary}

\begin{IEEEproof}
Define the vector $\mathbf{p}^{(\mathrm{FA})}=(P_{\mathrm{fa},1},\ldots,P_{\mathrm{fa},M})$. The PGF in \eqref{eq:e_fa} can be rewritten as $G_{\mathrm{FA}}(z)=
\prod_{m=1}^{M} \big[1-P_{\mathrm{fa},m}(1-z)\big]$. Expanding the product using the ESP definition \eqref{eq:esp_def} yields $G_{\mathrm{FA}}(z)
=\sum_{k=0}^{M} (-1)^k \xi_k\!\big(\mathbf{p}^{(\mathrm{FA})}\big)(1-z)^k$. Then, applying the binomial expansion $(1-z)^k=\sum_{i=0}^{k}(-1)^i\binom{k}{i}z^i$, we obtain
\begin{align}
G_{\mathrm{FA}}(z)
&=\sum_{k=0}^{M} (-1)^k \xi_k\big(\mathbf{p}^{(\mathrm{FA})}\big) \sum_{i=0}^{k}(-1)^i \binom{k}{i} z^i \nonumber\\
&= \sum_{i=0}^{M} \left[\sum_{k=i}^{M} (-1)^{k+i} \binom{k}{i} \xi_k\big(\mathbf{p}^{(\mathrm{FA})}\big)
\right] z^i \nonumber\\
&=\sum_{k=i}^{M}(-1)^{k+i}  \binom{k}{i} \xi_k\big(\mathbf{p}^{(\mathrm{FA})}\big) z^i.
\label{eq:pgf_fa_esp_step2}
\end{align}
Comparing the coefficients of $z^i$ in \eqref{eq:pgf_fa_esp_step2} gives \eqref{eq:ei_fa_esp}. Since the proof of \ref{eq:ei_d_esp} is similar with \ref{eq:ei_fa_esp}, we omit here.
\end{IEEEproof}


\section{System-Level DEP Derivation}
\label{sec:system_DEP}

In this section, we first analyze how the detection threshold partitions the domain into several key intervals determined by the wardens' breakpoints, and then exploit this piecewise structure to obtain the DEP expressions on each interval.

\subsection{Threshold Interval Structure of the DEP}

From (25)–(26), we observe that for each warden $m$, $P_{\text{fa},m}(\tau)$ and $P_{\text{md},m}(\tau)$ are piecewise–affine functions of the detection threshold $\tau$ (i.e., they are linear on each interval between breakpoints). The breakpoints are determined by the noise power and the channel–dependent parameters $\alpha_{1,m},\alpha_{2,m},\alpha_{3,m}$: when $\tau$ crosses these values, the functional form of $P_{\text{fa},m}$ or $P_{\text{md},m}$ switches between a constant and a linear term. In particular, $P_{\text{fa},m}$ changes at $\sigma_w^2$ and $\alpha_{1,m}$, while $P_{\text{md},m}$ changes at $\alpha_{2,m}$ and $\alpha_{3,m}$.

To describe the global behavior of all wardens, it is convenient to collect the corresponding extrema into seven key thresholds $\sigma^2,
  \ \alpha_1^{\min}, \alpha_2^{\min}, \alpha_3^{\min}, \alpha_1^{\max}, \alpha_2^{\max}, \alpha_3^{\max}$, where $\alpha_\ell^{\min} \triangleq \min_m \alpha_{\ell,m}$ and $\alpha_\ell^{\max} \triangleq \max_m \alpha_{\ell,m}$ for $\ell \in \{1,2,3\}$. These thresholds jointly determine the piecewise structure of the system–level false–alarm and miss–detection probabilities under majority voting.

By construction, we always have $\sigma^2 < \alpha_\ell^{\min} \le \alpha_\ell^{\max}, \ell \in \{1,2,3\}$,  and $\alpha_\ell^{\min} < \alpha_3^{\max}$ for $\ell=1,2$. The only remaining uncertainty concerns a few relative orders such as $(\alpha_1^{\min},\alpha_2^{\min})$ and $(\alpha_3^{\min},
\alpha_1^{\max})$. Instead of enumerating all $7 \choose 2$ pairwise comparisons, we classify the system behavior according to which type of local parameter reaches the probabilistic regime first as $\tau$ increases from $\sigma^2$. This leads to two representative
cases:
\begin{itemize}
  \item \textbf{ $\alpha_2^{\min} \le \alpha_1^{\min}$}.
  In this case, some wardens begin to exhibit probabilistic
  miss–detection before any false–alarm probability changes.
  \item \textbf{$\alpha_2^{\min} > \alpha_1^{\min}$}.
  Here, some wardens first enter a probabilistic false–alarm
  region while all miss–detection probabilities remain
  deterministic.
\end{itemize}
Although the larger thresholds (e.g., $\alpha_\ell^{\max}$ and $\alpha_3^{\min}$) further partition the domain of $\tau$, they do
not change this basic ordering of events.

\begin{figure}[t]
  \centering
  \begin{tikzpicture}[scale=0.9,>=stealth]
    \draw[->] (0,0) -- (7.2,0) node[right] {$\tau$};
    \draw[->] (0,0) -- (0,3.2) node[above] {$1$};

    \draw[dashed,red!70] (1.2,0) -- (1.2,3.2) node[above] {\scriptsize $\alpha_{1,1}$};
    \draw[dashed,red!70] (3.0,0) -- (3.0,3.2) node[above] {\scriptsize $\alpha_{2,1}$};
    \draw[dashed,red!70] (5.0,0) -- (5.0,3.2) node[above] {\scriptsize $\alpha_{3,1}$};

    \draw[dashed,blue!70] (1.8,0) -- (1.8,3.2) node[above] {\scriptsize $\alpha_{1,2}$};
    \draw[dashed,blue!70] (3.8,0) -- (3.8,3.2) node[above] {\scriptsize $\alpha_{2,2}$};
    \draw[dashed,blue!70] (6.0,0) -- (6.0,3.2) node[above] {\scriptsize $\alpha_{3,2}$};

    \draw[very thick,red]
      (0,2.9) -- (1.2,2.9)        
      -- (5.0,1.1)                
      -- (7.0,0.9);               

    \draw[very thick,red,dashed]
      (0,0.2) -- (3.0,0.2)        
      -- (5.0,1.9)                
      -- (7.0,2.8);               

    \draw[very thick,blue]
      (0,2.6) -- (1.8,2.6)        
      -- (6.0,1.0)                
      -- (7.0,0.85);              

    \draw[very thick,blue,dashed]
      (0,0.35) -- (3.8,0.35)      
      -- (6.0,2.1)                
      -- (7.0,2.9);               

    \node at (0.8,0.7) {\small phase 1};
    \node at (3.0,2.4) {\small phase 2};
    \node at (6.3,2.4) {\small phase 3};

    \begin{scope}[shift={(4.2,0.4)}]
      \draw[very thick,red] (0,2.4) -- (0.8,2.4);
      \node[right] at (0.9,2.4) {\scriptsize Warden 1, $P_{\text{ma},1}$};
      \draw[very thick,red,dashed] (0,2.0) -- (0.8,2.0);
      \node[right] at (0.9,2.0) {\scriptsize Warden 1, $P_{\text{md},1}$};

      \draw[very thick,blue] (0,1.6) -- (0.8,1.6);
      \node[right] at (0.9,1.6) {\scriptsize Warden 2, $P_{\text{fa},2}$};
      \draw[very thick,blue,dashed] (0,1.2) -- (0.8,1.2);
      \node[right] at (0.9,1.2) {\scriptsize Warden 2, $P_{\text{md},2}$};
    \end{scope}

  \end{tikzpicture}
  \caption{Qualitative evolution of the local false-alarm and miss-detection probabilities of two different wardens as the global threshold $\tau$ varies.}
  \label{fig:two_wardens}
\end{figure}
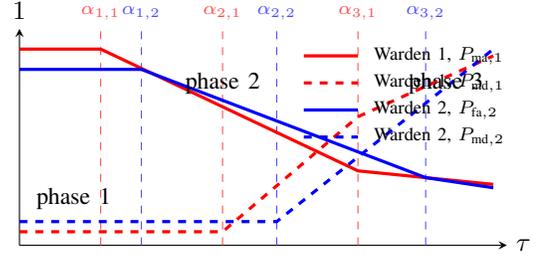

From Fig.~\ref{fig:two_wardens}, it can be observed that even within the same interval of $\tau$, the values of $P_{fa,m}$ and $P_{md,m}$ for different wardens are generally different, due to their distinct parameter triplets $(\alpha_{1,m},\alpha_{2,m},\alpha_{3,m})$. As a result, the system DEP under majority voting cannot be described by a single pair of local error probabilities; instead, we must keep track of \emph{how many} wardens have already crossed each type
of breakpoint. To make the subsequent analysis more transparent, we therefore reorder the three parameter sets in nondecreasing
order: $\{\alpha_{1,(l)}\}_{l=1}^M:\alpha_{1,(1)} \le \cdots \le \alpha_{1,(M)}, \{\alpha_{2,(k)}\}_{k=1}^M: \alpha_{2,(1)} \le \cdots \le \alpha_{2,(M)},\{\alpha_{3,(s)}\}_{s=1}^M:\alpha_{3,(1)} \le \cdots \le \alpha_{3,(M)}$. So, the global extrema can then be written as
$\alpha_\ell^{\min} = \alpha_{\ell,(1)}$ and $\alpha_\ell^{\max} = \alpha_{\ell,(M)}$ for $\ell\in\{1,2,3\}$.
For any given threshold $\tau$, we define the index functions
\begin{align}
  l &\triangleq
    \max\{l \in \{1,\ldots,M\}: \alpha_{1,(l)} \le \tau\,\},\\
  k &\triangleq
    \max\{k \in \{1,\ldots,M\}: \alpha_{2,(k)} \le \tau\,\},\\
  s &\triangleq
    \max\{s \in \{1,\ldots,M\}: \alpha_{3,(s)} \le \tau\,\}.
\end{align}
Intuitively, $l$, $k$, and $s$ count how many wardens have already crossed the corresponding breakpoints at
threshold $\tau$. As $\tau$ increases, these indices are nondecreasing step functions.

It is also useful to introduce the interval notation $I_l \triangleq [\alpha_{1,(l)}, \alpha_{1,(l+1)}),
  L_k \triangleq [\alpha_{2,(k)}, \alpha_{2,(k+1)}),
  K_s \triangleq [\alpha_{3,(s)}, \alpha_{3,(s+1)})$,
where we set $\alpha_{\ell,(M+1)} = +\infty$. On any fixed product interval $I_l \cap L_k \cap K_s$, the vector of all local
error probabilities $\big(P_{\text{fa},1}(\tau),\ldots,P_{\text{fa},M}(\tau), P_{\text{md},1}(\tau),\ldots,P_{\text{md},M}(\tau)\big)$ is an affine function of $\tau$ whose structure is completely determined by $(l,k,s)$.

\subsection{Closed-Form of the System's DEP}

To facilitate the subsequent derivations, we define the normalized constants $\alpha_m=\frac{\alpha_{1,m}}{P_J^{\max}A_J}$, $c_m=\frac{\alpha_{2,m}}{P_J^{\max}A_J}$ and $b=\frac{1}{P_J^{\max}A_J}$, so that, in the linear regimes $P_{\mathrm{fa},m}(\tau)=\alpha_m-b\tau$  $P_{\mathrm{md},m}(\tau)=b\tau-c_m$. With this normalization, the dependence on $\tau$ is captured entirely by the common slope $b$, which greatly simplifies the polynomial expressions appearing in the ESP/PGF-based analysis.

Building on the PGF representation in Theorem~\ref{thm:majority_dep_pgf} and the ESP-based coefficients in Corollary~\ref{coro:ei}, we next track how the triple $(l,k,s)$ of active index-set sizes evolves with $\tau$ for the two cases $\alpha_2^{\min} \le \alpha_1^{\min}$ and $\alpha_2^{\min} > \alpha_1^{\min}$. On each resulting threshold interval, we substitute the corresponding $\{P_{\mathrm{fa},m}(\tau), P_{\mathrm{md},m}(\tau)\}$ into the PGF expressions to obtain closed-form representations of $P_{\mathrm{FA}}(\tau)$ and $P_{\mathrm{MD}}(\tau)$, and hence the system DEP.

\begin{theorem}
\label{thm:dep_piecewise_full}
Consider the PASS-enhanced covert communication system with $M$ wardens employing the majority-voting fusion rule. If the ordered thresholds satisfy $\alpha_2^{\min}\le\alpha_1^{\min}$, then the detection error probability $P_{\mathrm{dep}}(\tau)$ is determined as \eqref{eq:dep_piecewise_full}
\begin{figure*}[!t]
\begin{equation}
P_{\mathrm{dep}}(\tau)=
\begin{cases}
1, & 0 \le \tau \le \sigma^2, \tau \ge \alpha_3^{\max}\\
\sum_{k=T}^{M}(-1)^{k+T}\binom{k-1}{T-1}\psi_{\tau}(k,M), & \sigma^2 < \tau \le \alpha_2^{\min}, \\
\sum_{k=T}^{M}(-1)^{k+T}\binom{k-1}{T-1}\psi_{\tau}(k,M) + \Psi_{\tau}(M,\psi_{\tau}(l,k)),   & \tau\in L_k\cap[\alpha_2^{\min},\alpha_1^{\min}], \\
\Phi_{\tau}(|J|,\psi_{\tau}(k,|J|)+ \Psi_{\tau}(M,\psi(l,k)),   & \tau\in I_l\cap L_k\cap[\alpha_1^{\min},\min(\alpha_1^{\max} \alpha_3^{\min})] \\
 \Phi_{\tau}(|J|,\psi_{\tau}(k,|J|)+ \Psi_{\tau}(M-s,\psi_{\tau}(l,k-s)),  & \tau\in I_l\cap L_k\cap K_s\cap[\alpha_3^{\min},\alpha_1^{\max}],\\
\Psi_{\tau}(M-s,\psi_{\tau}(l,k-s)),        & \tau\in L_k\cap K_s\cap[\alpha_1^{\max},\alpha_3^{\max}],
\end{cases}
\label{eq:dep_piecewise_full}
\end{equation}
\hrulefill
\end{figure*}
where the auxiliary polynomials $\psi_{\tau}(\cdot,\cdot)$, $\Phi_{\tau}(\cdot,\cdot)$ and $\Psi_{\tau}(\cdot,\cdot)$ are defined as
\begin{align}
\psi_{\tau}(x,y)
&=\sum_{r=0}^{x}(-b\tau)^r\binom{y-x+r}{r}\,\xi_{x-r}\big(\mathbf{a}_{(y)}\big), \\
\Phi_{\tau}(x,y)
&=\sum_{k=T}^{x}(-1)^{k+T}\binom{k-1}{T-1}
\sum_{r=0}^{k}(-b\tau)^r\binom{x-k+r}{r}\,y,  \\
\Psi_{\tau}(x,y)
&=\sum_{i=0}^{T-1}\sum_{k=i}^{x} (-1)^{k+i}\binom{k}{i}
\sum_{l=0}^{k}\binom{x-k}{k-l}\,y,
\end{align}
and $\xi_{m}(\cdot)$ denotes the $m$th elementary symmetric polynomial. The vector $\mathbf{a}_{(y)}$ selects the appropriate set of affine coefficients according to
\begin{equation}\label{eq:ay}
 \mathbf{a}_{(y)}=
 \begin{cases}
   \mathbf{a}_M=(a_m)_{m\in \{1,\ldots,M\}}, & \mbox{if } y=M,\\[2pt]
   \mathbf{a}_D=(1+c_m)_{m\in L_k}, & \mbox{if } y=k,\\[2pt]
   \mathbf{a}_J=(a_m)_{m\in \{1,\ldots,M\}\setminus I_l}, & \mbox{if } y=|J|,\\[2pt]
   \mathbf{a}'_D=(1+c_m)_{m\in L_k\setminus G}, & \mbox{if } y=k-s,
 \end{cases}
\end{equation}
so that the four arguments $y\in\{M,k,|J|,k-s\}$ correspond, respectively, to the full set of wardens, the linear-MD set $L_k$, the active-FA set $J$, and the linear-but-not-saturated MD set $L_k\setminus G$.
\end{theorem}

%

\begin{IEEEproof}
The proof proceeds by partitioning the threshold $\tau$ into several regimes, characterized by the breakpoints $\sigma^2$, $\alpha_2^{\min}$, $\alpha_1^{\min}$,
$\alpha_1^{\max}$, $\alpha_3^{\min}$, and $\alpha_3^{\max}$. In each regime we specify the per-warden error probabilities $\{P_{\mathrm{fa},m}(\tau),P_{\mathrm{md},m}(\tau)\}$, then substitute them into Theorem~\ref{thm:majority_dep_pgf} and Corollary~\ref{coro:ei} to obtain the DEP.

\textbf{1) When $\bm{\tau \in [0, \sigma^2]}$:}
In this interval, each warden operates in a deterministic regime with $P_{\mathrm{fa},m} = 1$ and $P_{\mathrm{md},m} = 0$ for all $m$. Substituting these values into the coefficient definitions in Eqs.~\eqref{eq:e_fa} and~\eqref{eq:e_d} yields
\begin{align}
e_i^{(\mathrm{FA})} = e_i^{(\mathrm{D})} =
\begin{cases}
1, & i = M,\\
0, & i \neq M.
\end{cases}
\end{align}
Hence, the system false-alarm and missed-detection probabilities satisfy $P_{\mathrm{FA}} = 1$ and $P_{\mathrm{MD}} = 0$, so the DEP is
$P_{\mathrm{dep}} = P_{\mathrm{FA}} + P_{\mathrm{MD}} = 1$.

\textbf{2) When $\bm{\tau\in[\sigma^2,\alpha_2^{\min}]}$:}
In this interval, every warden operates in the perfect-detection regime.
Because $\tau \le \alpha_2^{\min}$, each warden always detects the presence of the signal
whenever it exists, i.e., $P_{\mathrm{md},m}(\tau)=0$ for $m=1,\ldots,M$.
Meanwhile, the false-alarm probability decreases linearly with $\tau$ as $P_{\mathrm{fa},m}(\tau) =\alpha_m-b\tau$,
where $\alpha_m$ and $b$ are the normalized coefficients defined before the theorem.

Since all missed detections vanish, the DEP reduces to the majority-rule false-alarm tail.
By Theorem~\ref{thm:majority_dep_pgf},
\begin{equation}
P_{\mathrm{dep}}(\tau)
\!= \!P_{\mathrm{FA}}(\tau)
\!=\!\sum_{i=T}^{M} \!e^{(\mathrm{FA})}_i \big(P_{\mathrm{fa},1}(\tau),\ldots,P_{\mathrm{fa},M}(\tau)\big),
\end{equation}
where $e^{(\mathrm{FA})}_i$ is the ESP-based coefficient in~\eqref{eq:ei_fa_esp}.
To obtain a closed form, we invoke the ESP common-shift identity $\xi_k(x_1-c,\ldots,x_M-c)
=\sum_{r=0}^{k}(-c)^r \binom{M-k+r}{r}\,\xi_{k-r}(x_1,\ldots,x_M)$ with $x_m=a_m$ and $c=b\tau$.
Using the ESP expansion of $e^{(\mathrm{FA})}_i$ from Corollary~\ref{coro:ei}, we obtain
\begin{align}
P_{\mathrm{dep}}(\tau)
&=P_{\mathrm{FA}}(\tau)
=\sum_{i=T}^{M}\sum_{k=i}^{M} (-1)^{k+i}\binom{k}{i}\psi_{\tau}(k,M).
\label{eq:case2_intermediate}
\end{align}
Summing over $i$ and simplifying binomial coefficients yields
\begin{equation}
P_{\mathrm{dep}}(\tau)
=\sum_{k=T}^{M}(-1)^{k+T}\binom{k-1}{T-1}\psi_{\tau}(k,M).
\label{eq:dep_case2_poly}
\end{equation}

\textbf{3) When $\bm{\tau \in L_k \cap [\alpha_2^{\min}, \alpha_1^{\min}]}$:}
In this interval, $\tau < \alpha_1^{\min}$, so \emph{all} wardens remain in the linear false-alarm regime.
Consequently, the system false-alarm probability $P_{\mathrm{FA}}(\tau)$ retains the form in~\eqref{eq:dep_case2_poly}.
Meanwhile, exactly $k$ wardens enter the linear missed-detection regime as defined by $L_k$,
while the remaining $M-k$ wardens still have zero missed-detection probability:
\begin{equation}
P_{\mathrm{md},m}(\tau)=
\begin{cases}
b\tau-c_m, & m\in T_1,\\[2pt]
0, & m\in T_0,
\end{cases}
\end{equation}
where $T_1=\{m:\alpha_{2,m}\le\tau\}$ with $|T_1|=k$, and $T_0=\{m:\alpha_{2,m}>\tau\}$ with $|T_0|=M-k$.

To unify notation, define $\mathbf{q}(\tau)\triangleq (1-P_{\mathrm{md},1},\ldots,1-P_{\mathrm{md},M})$. By Theorem~\ref{thm:majority_dep_pgf} and Corollary~\ref{coro:ei},
\begin{align}
\label{eq:case3_pmd_reindex}
P_{\mathrm{MD}}(\tau)
&=\sum_{i=0}^{T-1}e_i^{(\mathrm{D})}\mathbf{q}(\tau)=\sum_{i=0}^{T-1} \sum_{k=i}^{M} (-1)^{k+i} \binom{k}{i}\,
\xi_k\!\big(\mathbf{q}(\tau) \big),
\end{align}
where $\xi_k(\cdot)$ denotes the $k$th ESP. Without loss of generality, reorder indices so that $T_1=\{1,\dots,k\}$ and $T_0=\{k+1,\ldots,M\}$. The $k$ active components of $\mathbf{q}(\tau)$ can be written as $\mathbf{q}(\tau)
\triangleq (1-P_{\mathrm{md},m}(\tau))_{m\in T_1}
=\mathbf{a}_{D}-b\tau\mathbf{1}_k$, where $\mathbf{a}_{D}\triangleq (1+c_m)_{m\in T_1}$ and $\mathbf{1}_k$ denotes the $k$-dimensional all-ones vector.

Since the remaining $M-k$ entries of $\mathbf{q}(\tau)$ are equal to one, their contribution can be separated using the ESP identity
\begin{align}\label{eq:q}
\xi_k(1\!-\!P_{\mathrm{md},1},\ldots,1\!-\!P_{\mathrm{md},M})
=\!\sum_{l=0}^{k}\! \binom{M\!-\!k}{k-l}\xi_l(\mathbf{q}(\tau)).
\end{align}
Applying again the common-shift identity to $\xi_l(\mathbf{q}(\tau))$ gives
\begin{equation}
\xi_l(\mathbf{q}(\tau))\!=\!\psi_{\tau}(l,k)
\!=\!\sum_{r=0}^{l}(-b\tau)^r\binom{k-l+r}{r}\xi_{l-r}(\mathbf{a}_{D}).
\end{equation}
Substituting (\ref{eq:q}) into (\ref{eq:case3_pmd_reindex}) and simplifying yields the closed-form expression of the system missed-detection probability:
\begin{align}\label{eq:case3_pmd}
P_{\mathrm{MD}}(\tau)
&=\sum_{i=0}^{T-1}\sum_{k=i}^{M} (-1)^{k+i}\binom{k}{i} \sum_{l=0}^{k}\binom{M-k}{k-l}\,\xi_l(\mathbf{q}(\tau)) \nonumber \\
&=\Psi_{\tau}(M,\psi_{\tau}(l,k)).
\end{align}
Combining~\eqref{eq:dep_case2_poly} and~\eqref{eq:case3_pmd}, we obtain
\begin{equation}
P_{\mathrm{dep}}(\tau)
=P_{\mathrm{FA}}(\tau)+P_{\mathrm{MD}}(\tau)
=\eqref{eq:dep_case2_poly}+\eqref{eq:case3_pmd}.
\label{eq:case3_dep_poly}
\end{equation}
which corresponds to the third row of~\eqref{eq:dep_piecewise_full}.

\textbf{4) When $\bm{\tau\in[\alpha_1^{\min},\alpha_1^{\max}]}$:} When $\tau$ enters $[\alpha_1^{\min},\alpha_1^{\max}]$, the local error probabilities $\{P_{\mathrm{fa},m}(\tau), P_{\mathrm{md},m}(\tau)\}$ no longer share a uniform structure across wardens: some nodes have already left the probabilistic false-alarm regime (so $P_{\mathrm{fa},m}(\tau)=0$), others still exhibit linear false-alarm behavior, and the missed-detection probabilities may start approaching the saturated value $1$. As illustrated in Figs.~\ref{fig:interval_intersections1} and~\ref{fig:interval_intersections2}, this is conveniently captured by the index sets $I_l$, $L_k$, and $K_s$, which remain constant within each intersection cell.

\begin{figure}[t]
  \centering
  \resizebox{0.95\columnwidth}{!}{
  \begin{tikzpicture}[>=Stealth,font=\footnotesize]
    \draw[->,ultra thick] (0,0) -- (12.5,0) node[below right] {$\tau$};

    \draw[thick] (2,0.1)--(2,-0.1) node[below] {$\bm{\alpha_1^{\min}}$};
    \draw[thick] (9,0.1)--(9,-0.1) node[below] {$\bm{\alpha_1^{\max}}$};
    \draw[->,dash dot,thick] (10.0,0.35)--(12.2,0.35);
    \node[anchor=west] at (10.1,0.55) {\scriptsize$\alpha_3^{\min}>\alpha_1^{\max}$};

    \fill[blue!10] (2,0.6) rectangle (9,1.4);
    \draw[thick] (2,0.6) rectangle (9,1.4);
    \node[anchor=west] at (4.1,1.05) {\small $I_l\!\cap\!L_k\!\cap\![\alpha_1^{\min},\alpha_1^{\max}]$};

    \foreach \x/\lab in {3.2/{\scriptsize$\alpha_{1,(l)}$}, 5.2/{\scriptsize$\alpha_{1,(l+1)}$}, 7.6/{\scriptsize$\alpha_{1,(l')}$}}{
      \draw[dashed,thick] (\x, -0.3) -- (\x, 1.7);
      \node[rotate=90] at (\x,-0.8) {\lab};
    }

    \foreach \x/\lab in {1.4/{\scriptsize$\alpha_{2,(k)}$}, 4.0/{\scriptsize$\alpha_{2,(k+1)}$}, 6.6/{\scriptsize$\alpha_{2,(k')}$}}{
      \draw[densely dotted,thick] (\x, -0.3) -- (\x, 1.7);
      \node[rotate=90] at (\x,-0.8) {\lab};
    }

    \draw[dash dot,thick] (11.5,-0.3)--(11.5,1.7);
    \node[rotate=90] at (11.5,-0.8) {\scriptsize$\alpha_{3,(s)}$};

    \draw[rounded corners=2pt,thick,black!70] (2.2,1.8) rectangle (8.8,3.25);
    \node[anchor=west] at (2.35,2.9) {\textbf{\scriptsize In $[\alpha_1^{\min},\alpha_1^{\max}]$ with $\alpha_3^{\min}>\alpha_1^{\max}$:}};
    \node[anchor=west] at (2.35,2.6) {\scriptsize $m\in I_l$: $P_{\mathrm{fa},m}=0$;\quad $m\notin I_l$: $P_{\mathrm{fa},m}=a_m-b\tau$};
    \node[anchor=west] at (2.35,2.3) {\scriptsize $m\in L_k$: $P_{\mathrm{md},m}=b\tau-c_m$;\quad $m\notin L_k$: $P_{\mathrm{md},m}=0$};
    \node[anchor=west] at (2.35,2.05) {\scriptsize $K_s=\varnothing$ (no saturation occurs).};

  \end{tikzpicture}
  }
  \vspace{2pt}
   \caption{Threshold structure for $\alpha_3^{\min}>\alpha_1^{\max}$. Throughout $\tau\in [\alpha_1^{\min},\alpha_1^{\max}]$, no warden enters the saturated MD regime ($K_s=\varnothing$). so only the sets $I_l$ and $L_k$ remain active. Consequently, $P_{\mathrm{fa},m}\in\{0,a_m-b\tau\}$ and $P_{\mathrm{md},m}\in\{0,b\tau-c_m\}$, hold uniformly over the interval, enabling closed-form evaluation of $P_{dep}(\tau)$ on $I_l\cap L_k\cap[\alpha_1^{\min},\alpha_1^{\max}]$.}
  \label{fig:interval_intersections1}
\end{figure}
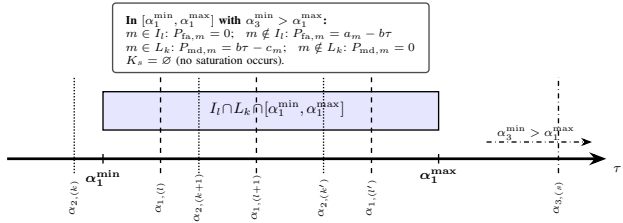

\begin{figure}[t]
  \centering
  \resizebox{0.95\columnwidth}{!}{
  \begin{tikzpicture}[>=Stealth,font=\footnotesize]
    \draw[->,ultra thick] (0,0) -- (12.5,0) node[below right] {$\tau$};

    \draw[thick] (2,0.1)--(2,-0.1) node[below] {$\bm{\alpha_1^{\min}}$};
    \draw[thick] (7,0.1)--(7,-0.1) node[below] {$\bm{\alpha_3^{\min}}$};
    \draw[thick] (11,0.1)--(11,-0.1) node[below] {$\bm{\alpha_1^{\max}}$};

    \fill[blue!12] (2,0.5) rectangle (7,1.3);
    \draw[thick] (2,0.5) rectangle (7,1.3);
    \node[anchor=west] at (3.1,0.9) {$I_l\!\cap\!L_k\!\cap\![\alpha_1^{\min},\alpha_3^{\min}]$};

    \fill[green!20] (7,0.5) rectangle (11,1.3);
    \draw[thick] (7,0.5) rectangle (11,1.3);
    \node[anchor=west] at (7.1,0.9) {$I_l\!\cap\!L_k\!\cap\!K_s\!\cap\![\alpha_3^{\min},\alpha_1^{\max}]$};

    \foreach \x/\lab in {2.8/{\scriptsize$\alpha_{1,(l)}$}, 4.5/{\scriptsize$\alpha_{1,(l+1)}$}, 9.2/{\scriptsize$\alpha_{1,(l')}$}}{
      \draw[densely dashed,thick] (\x,-0.3) -- (\x,1.6);
      \node[rotate=90] at (\x,-0.8) {\lab};
    }
    \foreach \x/\lab in {1.4/{\scriptsize$\alpha_{2,(k)}$}, 3.6/{\scriptsize$\alpha_{2,(k+1)}$}, 6.2/{\scriptsize$\alpha_{2,(k')}$}}{
      \draw[densely dotted,thick] (\x,-0.3) -- (\x,1.6);
      \node[rotate=90] at (\x,-0.8) {\lab};
    }
    \foreach \x/\lab in {7.6/{\scriptsize$\alpha_{3,(s)}$}, 8.8/{\scriptsize$\alpha_{3,(s+1)}$}, 10.2/{\scriptsize$\alpha_{3,(s')}$}}{
      \draw[dash dot,thick] (\x,-0.3) -- (\x,1.6);
      \node[rotate=90] at (\x,-0.8) {\lab};
    }

    \draw[rounded corners=2pt,black!70] (2.3,1.6) rectangle (6.5,3.4);
    \node[anchor=west] at (2.6,3.20) {\textbf{\scriptsize In $[\alpha_1^{\min},\alpha_3^{\min}]$:}};
    \node[anchor=west] at (2.6,2.80) {\scriptsize $m\in I_l$: $P_{\mathrm{fa},m}=0$};
    \node[anchor=west] at (2.6,2.47) {\scriptsize $m\notin I_l$: $P_{\mathrm{fa},m}=a_m-b\tau$};
    \node[anchor=west] at (2.56,2.13) {\scriptsize $m\in L_k$: $P_{\mathrm{md},m}=b\tau-c_m$};
    \node[anchor=west] at (2.56,1.80) {\scriptsize $m\notin L_k$: $P_{\mathrm{md},m}=0$};

    \draw[rounded corners=2pt,black!70] (7.1,1.6) rectangle (11,3.4);
    \node[anchor=west] at (7.2,3.20) {\textbf{\scriptsize In $[\alpha_3^{\min},\alpha_1^{\max}]$:}};
    \node[anchor=west] at (7.2,2.85) {\scriptsize $m\in I_l$: $P_{\mathrm{fa},m}=0$};
    \node[anchor=west] at (7.2,2.60) {\scriptsize $m\notin I_l$: $P_{\mathrm{fa},m}=a_m-b\tau$};
    \node[anchor=west] at (7.2,2.30) {\scriptsize $m\in K_s$: $P_{\mathrm{md},m}=1$};
    \node[anchor=west] at (7.2,2.05) {\scriptsize $m\in L_k\!\!\setminus \!\!K_s$: \!$P_{\mathrm{md},m}\!=b\tau-c_m$};
    \node[anchor=west] at (7.2,1.80) {\scriptsize $m\notin L_k$: $P_{\mathrm{md},m}=0$};

  \end{tikzpicture}
  }
  \vspace{2pt}
  \caption{Threshold structure for $\alpha_3^{\min}\le \alpha_1^{\max}$. Over $\tau\in[\alpha_1^{\min},\alpha_1^{\max}]$ (blue region), all wardens remain non-saturated ($K_s=\emptyset$), so only the sets $I_l$ and $L_k$ are active, and
  $P_{\mathrm{fa},m}\in\{0,a_m-b\tau\}$ and $P_{\mathrm{md},m}\in\{0,b\tau-c_m\}$. Over $\tau\in[\alpha_3^{\min},\alpha_1^{\max}]$ (green region), a subset of wardens enters the saturated MD regime, activating $K_s$ and producing  $P_{\mathrm{md},m}=1$ for $m\in K_s$. Within each region the index sets $(I_l,L_k,K_s)$ remain fixed, so the local error probabilities keep invariant forms (0/linear/1), enabling the closed-form evaluation of$P_{\mathrm{dep}}(\tau)$ on both subintervals.}
  \label{fig:interval_intersections2}
\end{figure}
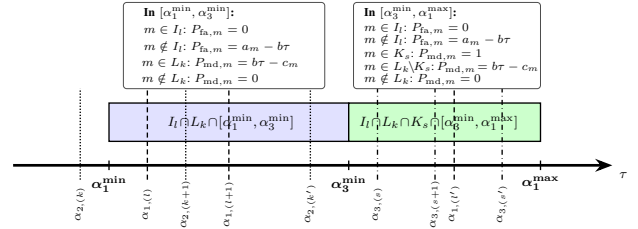

A key issue in this interval is whether $\tau$ exceeds the smallest saturation breakpoint $\alpha_3^{\min} \triangleq \min_m \alpha_{3,m}$.
If $\alpha_3^{\min} > \alpha_1^{\max}$, then no warden enters the full missed-detection regime on $[\alpha_1^{\min},\alpha_1^{\max}]$, so $K_s=\varnothing$.
If $\alpha_3^{\min} \le \alpha_1^{\max}$, there exists a sub-interval where $P_{\mathrm{md},m}(\tau)=1$ for some wardens, i.e., $K_s\neq\varnothing$. We therefore distinguish these two cases.

\textbf{Case A} $(\alpha_3^{\min}>\alpha_1^{\max})$: Here the smallest saturation breakpoint lies to the right of $[\alpha_1^{\min},\alpha_1^{\max}]$,
so $K_s=\varnothing$. As shown in Fig.~\ref{fig:interval_intersections1}, for all $\tau\in I_l\cap L_k\cap[\alpha_1^{\min},\alpha_1^{\max}]$,
\begin{equation}
P_{\mathrm{fa},m}\!=\!
\begin{cases}
0, & m\in I_l,\\[2pt]
a_m-b\tau, & m\notin I_l,
\end{cases}
P_{\mathrm{md},m}=
\begin{cases}
b\tau-c_m, & m\in L_k,\\[2pt]
0, & m\notin L_k,
\end{cases}
\label{eq:caseIlLk_piece}
\end{equation}
with $a_m$, $b$, and $c_m$ as defined above. Thus, every warden is either in a linear or zero regime, and this (0/linear) structure is invariant
within each cell $I_l\cap L_k$.

For the false-alarm part, only wardens in $J\triangleq\{1,\ldots,M\}\setminus I_l$ have nonzero false-alarm probabilities. Hence $P_{\mathrm{FA}}(\tau)$ is the majority-tail of a Poisson–binomial distribution with $|J|$ parameters $\{a_m-b\tau:m\in J\}$. Applying the derivation of~\eqref{eq:dep_case2_poly} to this active subset yields
\begin{align}
P_{\mathrm{FA}}(\tau)
&=\sum_{k=T}^{|J|}(-1)^{k+T}\binom{k-1}{T-1}\psi_{\tau}(k,|J|) \nonumber \\
&=\Phi_{\tau}(|J|,\psi_{\tau}(k,|J|)).
\label{eq:case41_pfa}
\end{align}

The missed-detection structure in this case is identical to that in $\tau\in L_k\cap[\alpha_2^{\min},\alpha_1^{\min}]$, so $P_{\mathrm{MD}}(\tau)$ is given by~\eqref{eq:case3_pmd}. Combining the two contributions, the DEP on $I_l\cap L_k\cap[\alpha_1^{\min},\alpha_1^{\max}]$ is
\begin{equation}\label{eq:dep-4}
P_{\mathrm{dep}}(\tau)
=P_{\mathrm{FA}}(\tau)+P_{\mathrm{MD}}(\tau)
=\eqref{eq:case41_pfa}+\eqref{eq:case3_pmd},
\end{equation}
which corresponds to the fourth row of~\eqref{eq:dep_piecewise_full} with $K_s=\varnothing$.

\textbf{Case B} $(\alpha_3^{\min}\le\alpha_1^{\max})$: In this case, some wardens enter the full missed-detection regime once $\tau$ exceeds $\alpha_3^{\min}$.
As illustrated in Fig.~\ref{fig:interval_intersections2}, the interval
$[\alpha_1^{\min},\alpha_1^{\max}]$ is split as
\begin{equation}
[\alpha_1^{\min},\alpha_1^{\max}]=\underbrace{[\alpha_1^{\min},\alpha_3^{\min}]}_{\text{no full MD: }K_s=\varnothing}
\cup
\underbrace{[\alpha_3^{\min},\alpha_1^{\max}]}_{\text{with full MD: }K_s\neq\varnothing}.
\end{equation}
On each cell $I_l\cap L_k\cap[\alpha_1^{\min},\alpha_3^{\min}]$, we still have $K_s=\varnothing$,
so $\{P_{\mathrm{fa},m},P_{\mathrm{md},m}\}\in\{0,\text{linear}\}$ and the local structure is exactly
the same as in Case~A.
Thus, the DEP on $I_l\cap L_k\cap[\alpha_1^{\min},\alpha_3^{\min}]$ is given by~\eqref{eq:dep-4}.

On the partial-saturation cells
$I_l\cap L_k\cap K_s\cap[\alpha_3^{\min},\alpha_1^{\max}]$, the missed-detection probabilities take three possible forms:
\begin{equation}
P_{\mathrm{md},m}(\tau)=
\begin{cases}
0,& m\in T_0,\\
b\tau-c_m,& m\in T_1',\\
1,& m\in G,
\end{cases}
\end{equation}
where $T_0=\{m:\tau<\alpha_{2,m}\}$, $T_1'=\{m:\alpha_{2,m}\le \tau\le \alpha_{3,m}\}$,
$G=\{m:\tau>\alpha_{3,m}\}$, and $|T_0|=M-k$, $|T_1'|=k-s$, $|G|=s$.
Wardens in $T_0$ have $P_{\mathrm{md},m}=0$, those in $T_1'$ have linear $P_{\mathrm{md},m}$, and those in
$G$ are fully saturated with $P_{\mathrm{md},m}=1$.

Define the active detection vector over $T_1'$ as $\mathbf{q}'(\tau) \triangleq (1-P_{\mathrm{md},m}(\tau))_{m\in T_1'}=\mathbf{a}'_{D}-b\tau\,\mathbf{1}_{k-s}$, where $\mathbf{a}'_{D}\triangleq (1+c_m)_{m\in T_1'}$.
The $M-k$ indices in $T_0$ contribute ones and the $s$ indices in $G$ contribute zeros to the vector
$(1-P_{\mathrm{md},1},\ldots,1-P_{\mathrm{md},M})$, so the missed-detection distribution is equivalent
to that of an $(M-s)$-warden system with $k-s$ active components.
Repeating the ESP-based derivation of \eqref{eq:case3_pmd_reindex}-\eqref{eq:case3_pmd} with the substitutions $M\to M-s,k\to k-s,
\mathbf{q}(\tau) \to \mathbf{q}'(\tau)$,
gives
\begin{align}\label{eq:case42_pmd}
P_{\mathrm{MD}}(\tau)&=\!\sum_{i=0}^{T-1}\sum_{k=i}^{M-s} (-1)^{k+i}\binom{k}{i}\!\sum_{l=0}^{k}\!\binom{M-s-k}{k-l}\xi_l\big(\mathbf{q}'(\tau)\big), \nonumber \\
&=\Psi_{\tau}(M-s,\psi_{\tau}(l,k-s)).
\end{align}

The false-alarm part is unaffected by saturation and still satisfies~\eqref{eq:case41_pfa}.
Thus, for $\tau\in I_l\cap L_k\cap K_s\cap[\alpha_3^{\min},\alpha_1^{\max}]$,
\begin{equation}
P_{\mathrm{dep}}(\tau)
=
P_{\mathrm{FA}}(\tau)+P_{\mathrm{MD}}(\tau)
=
\eqref{eq:case41_pfa}+\eqref{eq:case42_pmd},
\label{eq:case42_dep}
\end{equation}
which gives the fifth row of~\eqref{eq:dep_piecewise_full}.

\textbf{5) When $\bm{\tau \in L_k \cap K_s \cap [\alpha_1^{\max}, \alpha_3^{\max}]}$:} Once $\tau\ge \alpha_1^{\max}$, all wardens have left the linear false-alarm regime, so $P_{\mathrm{fa},m}(\tau)=0$ for all $m$ and hence $P_{\mathrm{FA}}(\tau) = 0$. In this interval, the individual missed-detection probabilities preserve exactly the same piecewise structure as in $\tau \in I_l \cap L_k \cap K_s \cap [\alpha_3^{\min}, \alpha_1^{\max}]$: the partition into $T_0$, $T_1'$ and $G$ remains unchanged, and $P_{\mathrm{md},m}(\tau)=b\tau-c_m$ for all $m \in T_1'$ still holds. As a result, the PGF of the missed-detection distribution and the system $P_{\mathrm{MD}}(\tau)$ are identical in form to those in the partial-saturation regime. Therefore, $P_{\mathrm{MD}}(\tau)$ is given by the ESP-based expression in~\eqref{eq:case42_pmd}, and $P_{\mathrm{dep}}(\tau)=P_{\mathrm{MD}}(\tau)$, which corresponds to the last nontrivial row of~\eqref{eq:dep_piecewise_full}.

\textbf{6) When $\bm{\tau \in [\alpha_3^{\max}, +\infty)}$:} According to~\eqref{eq:fa} and~\eqref{eq:md}, each warden satisfies
$P_{\mathrm{FA},m}(\tau)=0$ and $P_{\mathrm{MD},m}(\tau)=1$. Hence, the global false-alarm probability is zero, whereas the missed-detection probability
equals one. Consequently, the system DEP in this regime is again $P_{\mathrm{dep}}(\tau)=1$.

Collecting the DEP expressions obtained in the above six regimes and aligning the corresponding threshold intervals yields the piecewise closed-form DEP in
\eqref{eq:dep_piecewise_full}, which completes the proof.
\end{IEEEproof}

When $\alpha_2^{\min}>\alpha_1^{\min}$, the evolution of the system DEP across the threshold $\tau$ follows the same structural pattern as in the case $\alpha_2^{\min}\le \alpha_1^{\min}$. The only difference lies in the ordering of the breakpoints and the resulting index sets. By adapting the same ESP-based framework, we obtain the following complete characterization.

\begin{theorem}
\label{thm:dep_piecewise_full1}
Consider the PASS-enhanced covert communication system with $M$ wardens employing the majority-voting fusion rule. If the ordered parameters satisfy $\alpha_2^{\min}>\alpha_1^{\min}$, then the detection error probability $P_{\mathrm{dep}}(\tau)$ is given by the piecewise expression in \eqref{eq:dep_piecewise_full1}:
\begin{figure*}[t]
\begin{equation}
P_{\mathrm{dep}}(\tau)=
\begin{cases}
1, & 0 \le \tau \le \sigma^2,\; \tau \ge \alpha_3^{\max},\\[2pt]
\sum_{k=T}^{M}(-1)^{k+T}\binom{k-1}{T-1}\psi_{\tau}(k,M),
& \sigma^2 < \tau \le \alpha_1^{\min}, \\[2pt]
\Phi_{\tau}(|J|,\psi_{\tau}(k,|J|)),
& \tau\in L_k\cap[\alpha_1^{\min},\alpha_2^{\min}], \\[2pt]
\Phi_{\tau}(|J|,\psi_{\tau}(k,|J|))+\Psi_{\tau}(M,\psi_{\tau}(l,k)),
& \tau\in I_l\cap L_k\cap[\alpha_2^{\min},\min(\alpha_1^{\max}, \alpha_3^{\min})],\\[2pt]
\Phi_{\tau}(|J|,\psi_{\tau}(k,|J|))+\Psi_{\tau}(M-s,\psi_{\tau}(l,k-s)),
& \tau\in I_l\cap L_k\cap K_s\cap[\alpha_3^{\min},\alpha_1^{\max}],\\[2pt]
\Psi_{\tau}(M-s,\psi_{\tau}(l,k-s)),
& \tau\in L_k\cap K_s\cap[\alpha_1^{\max},\alpha_3^{\max}].
\end{cases}
\label{eq:dep_piecewise_full1}
\end{equation}
\hrulefill
\end{figure*}
Here the auxiliary polynomials $\psi_{\tau}(\cdot,\cdot)$, $\Phi_{\tau}(\cdot,\cdot)$, $\Psi_{\tau}(\cdot,\cdot)$, and the vector $\mathbf{a}_{(y)}$ are the same as in Theorem~\ref{thm:dep_piecewise_full}.
\end{theorem}

\begin{IEEEproof}
The proof proceeds by determining the system DEP on each interval induced by the ordering $\sigma^2<\alpha_1^{\min}<\alpha_2^{\min}<\alpha_1^{\max}<\alpha_3^{\min}<\alpha_3^{\max}$. Three ranges require separate treatment: $\tau \in [\sigma^2,\alpha_1^{\min}], \tau \in I_l\cap[\alpha_1^{\min},\alpha_2^{\min}],\tau \in [\alpha_2^{\min},\alpha_1^{\max}]$.

when $\tau\in[\sigma^2,\alpha_1^{\min}]$, since $\alpha_2^{\min}>\alpha_1^{\min}$, every $\tau$ in this range also satisfies $\tau<\alpha_2^{\min}$.
Thus each warden behaves exactly as in the interval $[\sigma^2,\alpha_2^{\min}]$ of the case
$\alpha_2^{\min}>\alpha_1^{\min}$. Accordingly, the DEP is identical Eq.~\eqref{eq:dep_case2_poly}.

For $\tau\in I_l\cap[\alpha_1^{\min},\alpha_2^{\min}]$, because $\tau<\alpha_2^{\min}$ still holds, each warden remains in the perfect-detection regime, giving $P_{\mathrm{md},m}=0, m=1,\ldots,M$. For the false alarms, $P_{\mathrm{fa},m}(\tau)=0$ when $m\in I_l$, and $P_{\mathrm{fa},m}(\tau)=a_m-b\tau, m\notin I_l$.  Thus the system DEP equals the system false-alarm probability, and the resulting expression matches Eq.~(\ref{eq:case41_pfa}).

When $\tau\in[\alpha_2^{\min},\alpha_1^{\max}]$, in this range, the structure of the local error probabilities mirrors that in previous, except that the interval is shifted. For each fixed $\tau$, we construct the index sets $I_1(\tau)$, $L_k(\tau)$, and $K_s(\tau)$ exactly as in Subsection~V-B.
The vectors $\big(P_{\mathrm{fa},m}(\tau)\big)$ and $\big(1-P_{\mathrm{md},m}(\tau)\big)$ again take values in $\{0,\text{linear},1\}$, and hence the ESP-based formulas \eqref{eq:case3_pmd}, \eqref{eq:case41_pfa}, and \eqref{eq:case42_pmd} apply verbatim.
Only the admissible $\tau$-range and the composition of the sets change.

As before, the characterization depends on the relative ordering of $\alpha_3^{\min}$ and $\alpha_1^{\max}$. If $\alpha_3^{\min}>\alpha_1^{\max}$, no warden reaches the saturated missed-detection state; hence $K_s(\tau)=\varnothing$ on the entire interval.
The system remains in the non-saturated region, and $P_{\mathrm{dep}}(\tau)=\eqref{eq:case41_pfa}+\eqref{eq:case3_pmd}, \tau\in[\alpha_2^{\min},\alpha_1^{\max}]$. If $\alpha_3^{\min}\le\alpha_1^{\max}$, the region $[\alpha_2^{\min},\alpha_1^{\max}]$ splits at $\alpha_3^{\min}$.
For $\tau<\alpha_3^{\min}$, we again have $K_s(\tau)=\varnothing$, yielding $P_{\mathrm{dep}}(\tau)
=\eqref{eq:case41_pfa}+\eqref{eq:case3_pmd}$. For $\tau\ge\alpha_3^{\min}$, some wardens satisfy $P_{\mathrm{md},m}(\tau)=1$.
As in previous the saturated indices factor out of the ESP representation, giving the mixed-regime expression: $P_{\mathrm{dep}}(\tau)
=\eqref{eq:case41_pfa}+\eqref{eq:case42_pmd}$.

Putting these cases together yields exactly the piecewise form in \eqref{eq:dep_piecewise_full1}.
\end{IEEEproof}

\subsection{Problem Formulation and Optimal Design}

In this paper, we aim to jointly optimize the power allocation, power-radiation coefficients, and PA spatial configuration to maximize the average covert rate (ACR) at Bob under randomized jamming. Conveniently, let $\boldsymbol{\Psi} \triangleq \big\{P_C, P_J^{\max}, \mathbf{\Theta}_C^*, \mathbf{\Theta}_J^*, \boldsymbol{x}_C, \boldsymbol{x}_J\big\}$ collect all design variables. Since the wardens' detection thresholds are not under Alice's control, we adopt a conservative assumption that the wardens choose the detection threshold that minimizes the system-level DEP for a given configuration. This worst-case viewpoint captures the adversarial nature of covert communications and leads to a robust design that guarantees a lower bound on the achievable covert rate. Accordingly, the overall optimization problem is formulated as
\begin{subequations}
\label{eq:optimization}
\begin{align}
&\max_{\boldsymbol{\Psi}} \quad
\bar R_C(\boldsymbol{\Psi})
= \mathbb{E}_{ P_J}\!\left[ \log_2(1+\gamma_b(\boldsymbol{\Psi} ))\right],
\label{eq:obj} \\
&\quad\text{s.t.}\quad \min_{\tau}\,\, P_{\text{dep}}(\tau,\boldsymbol{\Psi}) \geq 1 - \epsilon, \label{eq:const1} \\
&\quad\quad\quad 0 \leq P_C,\ \ 0 \leq P_J^{\max},\ \ P_C + P_J^{\max} \leq P_{\text{max}}, \label{eq:const2}\\
&\quad\quad\quad\quad\quad 0 \leq x^C_{n_1} \leq L,\ \ 0 \leq x^J_{n_2} \leq L, \label{eq:const5}\\
&\quad\quad\quad\quad\quad x^C_{n_1+1} - x^C_{n_1} \geq \Delta x_{\text{min}}, \label{eq:const6}\\
&\quad\quad\quad\quad\quad x^J_{n_2+1} - x^J_{n_2} \geq \Delta x_{\text{min}}. \label{eq:const7}
\end{align}
\end{subequations}
Constraint~\eqref{eq:const1} enforces the system-level covertness requirement in a worst-case sense by ensuring that even under the optimal threshold selection at the wardens, the resulting DEP is no smaller than $1-\epsilon$, where $\epsilon\in(0,1)$ is a prescribed covertness parameter. Constraint~\eqref{eq:const2} limits the power budgets under randomized jamming: $P_J^{\max}$ specifies the randomization range of $P_J$, and the peak budget $P_C+P_J^{\max}\le P_{\max}$ guarantees the total transmit power does not exceed $P_{\max}$ for all realizations of $P_J$. Constraints~\eqref{eq:const5}--\eqref{eq:const7} ensure that all PAs are deployed within the physical waveguide of length $L$ and that the minimum element spacing is no smaller than $\Delta x_{\min}$, which can be chosen according to fabrication limits (typically $0.1$--$0.2$ times the guided wavelength $\lambda_g$).

In the follows, we develop an iterative solver that converts \eqref{eq:optimization} into a sequence of tractable convex subproblems while preserving monotonic improvement of a valid surrogate objective.

\subsection{Solver Based on the MM--BCD--SCA Framework}
\label{subsec:solver_mm_bcd_sca}

Problem~\eqref{eq:optimization} is non-convex due to (i) the expectation-based average-rate objective under randomized jamming, (ii) the bilinear coupling between the transmit power/radiation coefficients and the PA locations, and (iii) the worst-case covertness constraint involving $\min_{\tau>0} P_{\mathrm{dep}}(\tau,\boldsymbol{\Psi})$. To obtain a tractable and convergent design procedure, we develop a solver within an MM--BCD--SCA framework. Specifically, we (i) approximate the expectation in $\bar R_C(\boldsymbol{\Psi})$ via a quadrature rule, (ii) construct a globally valid MM minorizer for each resulting log-SINR term, and (iii) apply a two-block BCD strategy that alternates between updating the power--radiation variables and updating the PA positions, where each block is handled using successive convex approximation (SCA). Meanwhile, the covertness constraint is treated using a Danskin-type affine surrogate of the DEP value function $g(\boldsymbol{\Psi}) \triangleq \min_{\tau>0} P_{\mathrm{dep}}(\tau,\boldsymbol{\Psi})$, which is refreshed at every iteration. The resulting algorithm yields a non-decreasing sequence of surrogate objective values and converges to a stationary point of the corresponding surrogate problem (and its SDR relaxation when applicable).

\subsubsection{Average-Rate Surrogate via Quadrature and MM}
\label{subsubsec:avg_rate_surrogate}

In this subsection, we approximate the expectation in $\bar R_C(\boldsymbol{\Psi})$ using a quadrature rule and then construct an MM minorizer for each log-SINR term. This leads to a concave surrogate objective that can be optimized efficiently within SCA. Let the instantaneous jamming power be $\tilde P_J=\xi P_J^{\max}$ with $\xi\sim\mathcal{U}[0,1]$, and define
\begin{align}
S(\boldsymbol{\Psi})
&\triangleq \left|\bm{\mathrm{h}}_{b}^T(\boldsymbol{x}_{C})\sqrt{P_{C}}\mathbf{\Theta}^*_C\boldsymbol{\omega}_{C}\right|^2,\\
I(\boldsymbol{\Psi})
&\triangleq \left|\bm{\mathrm{h}}_{b}^T(\boldsymbol{x}_{J})\sqrt{P_{J}^{\max}}\mathbf{\Theta}_J^*\boldsymbol{\omega}_{J} \right|^2,
\end{align}
so that the instantaneous SINR under realization $\xi$ is
\begin{equation}
\gamma_b(\xi;\boldsymbol{\Psi})=\frac{S(\boldsymbol{\Psi})}{\xi\, I(\boldsymbol{\Psi})+\sigma_b^2}.
\end{equation}
Accordingly, the average covert rate is $\bar R_C(\boldsymbol{\Psi})=\mathbb{E}_{\xi}\!\left[\log_2\!\left(1+\gamma_b(\xi;\boldsymbol{\Psi})\right)\right]$. To obtain a tractable form, we approximate the expectation by an $L$-point quadrature
\begin{equation}
\bar R_C(\boldsymbol{\Psi})
\approx \sum_{\ell=1}^{L} w_\ell\,
\log_2\!\left(1+\frac{S(\boldsymbol{\Psi})}{\xi_\ell I(\boldsymbol{\Psi})+\sigma_b^2}\right),
\label{eq:avg_rate_quadrature}
\end{equation}
where $\{\xi_\ell,w_\ell\}_{\ell=1}^L$ are fixed weights over $[0,1]$ (e.g., Gauss--Legendre), satisfying
$\xi_\ell\in[0,1]$, $w_\ell\ge 0$, and $\sum_{\ell=1}^L w_\ell=1$. For a given weight $\xi_\ell$, denote
\begin{align}
r_\ell(\boldsymbol{\Psi})
&\triangleq \log_2\left(1+\frac{S(\boldsymbol{\Psi})}{\xi_\ell I(\boldsymbol{\Psi})+\sigma_b^2}\right)\nonumber\\
&=\frac{1}{\ln 2}\Big[\ln\!\big(S(\boldsymbol{\Psi})+\xi_\ell I(\boldsymbol{\Psi})+\sigma_b^2\big)
-\ln\!\big(\xi_\ell I(\boldsymbol{\Psi})+\sigma_b^2\big)\Big].
\end{align}
Since $-\ln(\cdot)$ is convex, its first-order Taylor expansion gives a global under-estimator. Let $D_\ell^{(k)}\triangleq \xi_\ell I(\boldsymbol{\Psi}^{(k)})+\sigma_b^2$. Then, an MM minorizer of $r_\ell(\boldsymbol{\Psi})$ (tight at $\boldsymbol{\Psi}^{(k)}$) is given by
\begin{align}
r_\ell(\boldsymbol{\Psi})
\ge &\hat r_\ell(\boldsymbol{\Psi}|\boldsymbol{\Psi}^{(k)})
\triangleq \frac{1}{\ln 2}\Big[\ln\big(S(\boldsymbol{\Psi})+\xi_\ell I(\boldsymbol{\Psi})+\sigma_b^2\big) \nonumber\\
&-\ln\big(D_\ell^{(k)}\big)
-\frac{\xi_\ell}{D_\ell^{(k)}}\big(I(\boldsymbol{\Psi})-I(\boldsymbol{\Psi}^{(k)})\big)
\Big].
\label{eq:mm_lowerbound}
\end{align}
By construction, $\hat r_\ell(\boldsymbol{\Psi}|\boldsymbol{\Psi}^{(k)})$ satisfies \emph{tightness} and \emph{lower-bound} properties, i.e., it equals $r_\ell(\boldsymbol{\Psi})$ at $\boldsymbol{\Psi}^{(k)}$ and never exceeds $r_\ell(\boldsymbol{\Psi})$ elsewhere. Therefore, a concave surrogate of the average rate is given by
\begin{equation}
\hat{\bar R}_C(\boldsymbol{\Psi}\,|\,\boldsymbol{\Psi}^{(k)})
\triangleq \sum_{\ell=1}^{L} w_\ell\,\hat r_\ell(\boldsymbol{\Psi}\,|\,\boldsymbol{\Psi}^{(k)}),
\label{eq:mm_avg_rate}
\end{equation}
which can be maximized efficiently within each MM iteration.

\subsubsection{Covertness Handling via Danskin--SCA}

Define the DEP value function $g(\boldsymbol{\Psi})\triangleq \min_{\tau>0} P_{\mathrm{dep}}(\tau,\boldsymbol{\Psi})$, so that the covertness constraint is equivalently written as $g(\boldsymbol{\Psi})\ge 1-\epsilon$. For a given iterate $\boldsymbol{\Psi}^{(k)}$, we obtain an approximate minimizer $\tau^{(k)}$ via a one-dimensional search over a finite candidate set (e.g., breakpoints and stationary points) implied by the piecewise DEP expressions in Section~V. Motivated by Danskin’s theorem~\cite{mangasarian1994nonlinear}, we approximate a subgradient of $g(\boldsymbol{\Psi})$ at $\boldsymbol{\Psi}^{(k)}$ as
\begin{equation}
\mathbf{d}_k \triangleq
\nabla_{\boldsymbol{\Psi}} P_{\mathrm{dep}}(\tau,\boldsymbol{\Psi})
\big|_{\tau=\tau^{(k)},\,\boldsymbol{\Psi}=\boldsymbol{\Psi}^{(k)}}.
\end{equation}We then enforce the following affine approximation of the covertness constraint
\begin{equation}
g(\boldsymbol{\Psi}^{(k)}) + \mathbf{d}_k^T(\boldsymbol{\Psi}-\boldsymbol{\Psi}^{(k)}) \ge 1-\epsilon,
\label{eq:dep_linearized_new}
\end{equation}
which is refreshed at every iteration and imposed in the subsequent convexified subproblems. This linearization is used as an \emph{inner} (conservative) approximation in the SCA sense so that feasibility with respect to $g(\boldsymbol{\Psi})\ge 1-\epsilon$ is preserved in a neighborhood of $\boldsymbol{\Psi}^{(k)}$.

\subsubsection{Two-Block BCD Updates and Convex Subproblems}

We decouple the remaining variables using a two-block BCD scheme, where each block is solved from a convex surrogate constructed via SDR and SCA. To obtain tractable subproblems for the power/radiation variables, we adopt a lifted covariance representation
\begin{align}
\mathbf{V}_C &\triangleq \mathbf{v}_C \mathbf{v}_C^H,\quad
\mathbf{V}_J \triangleq \mathbf{v}_J \mathbf{v}_J^H. \\
\mathbf{H}_C(\boldsymbol{x}_C) &\triangleq \bm{\mathrm{h}}_b\bm{\mathrm{h}}_b^H,  \quad
\mathbf{H}_J(\boldsymbol{x}_J) \triangleq \bm{\mathrm{h}}_b\bm{\mathrm{h}}_b^H,
\end{align}
where $\mathbf{v}_C \triangleq \sqrt{P_{C}}\mathbf{\Theta}_C^* \boldsymbol{\omega}_{C}$, $\mathbf{v}_J \triangleq \sqrt{P_{J}}\mathbf{\Theta}_J^* \boldsymbol{\omega}_{J}$. Hence, we have
\begin{align}
  S(\boldsymbol{\Psi})
  = \mathrm{Tr}\big(\mathbf{H}_C\mathbf{V}_C\big), \quad
  I(\boldsymbol{\Psi})
  = \mathrm{Tr}\big(\mathbf{H}_J\mathbf{V}_J\big).
\end{align}
Here, the exact covariance matrices satisfy $\mathrm{rank}(\mathbf{V}_C)=\mathrm{rank}(\mathbf{V}_J)=1$ and
\begin{equation}
  \mathrm{diag}(\mathbf{V}_C) = P_{C}\mathbf{\Theta}_C^*,\quad
  \mathrm{diag}(\mathbf{V}_J) = P_{J}\mathbf{\Theta}_J^*.
  \label{eq:diag_constraints}
\end{equation}
Following the SDR idea, we relax the rank constraints and impose $\mathbf{V}_C\succeq \mathbf{0}$ and $\mathbf{V}_J\succeq \mathbf{0}$. To simplify notation and decouple diagonal constraints, we introduce auxiliary vectors
$\mathbf{s}_C\triangleq\mathrm{diag}(\mathbf{V}_C)$ and $\mathbf{s}_J\triangleq\mathrm{diag}(\mathbf{V}_J)$. Then, the concave MM surrogate of the average covert rate can be written as
\begin{align}
\!\!\!\!\hat{\bar R}_C(&\boldsymbol{\Psi}|\boldsymbol{\Psi}^{(k)})
\!=\!\!\sum_{\ell=1}^{L}\!\frac{w_\ell}{\ln 2}\!\Bigg[\!
\ln\!\Big(\mathrm{Tr}\big(\mathbf{H}_C\mathbf{V}_C\big)\!+\!\xi_\ell\mathrm{Tr}\big(\mathbf{H}_J\mathbf{V}_J\big)\!+\!\sigma_b^2\Big)
 \nonumber\\
&-\ln D_\ell^{(k)}-\frac{\xi_\ell}{D_\ell^{(k)}}\Big(\mathrm{Tr}\big(\mathbf{H}_J\mathbf{V}_J\big)\!-\!\mathrm{Tr}\big(\mathbf{H}_J\mathbf{V}_J^{(k)}\big)\Big)
\Bigg].
\label{eq:mm_avg_rate_trace}
\end{align}

For notational brevity, we define the lifted variable set $\tilde{\boldsymbol{\Psi}}\triangleq\{\mathbf{V}_C,\mathbf{V}_J,\mathbf{s}_C,\mathbf{s}_J,P_C,P_J^{\max},\boldsymbol{x}_C,\boldsymbol{x}_J\}$.
At MM iteration $k$, we solve the following surrogate problem
\begin{subequations}
\label{eq:mm_surrogate_problem}
\begin{align}
\max_{\tilde{\boldsymbol{\Psi}}}\quad
& \hat{\bar R}_C(\boldsymbol{\Psi}\,|\,\boldsymbol{\Psi}^{(k)})
\label{eq:mm_obj}\\
\text{s.t.}\quad
& \eqref{eq:dep_linearized_new},\;
\eqref{eq:diag_constraints},\;
\mathbf{V}_C\succeq \mathbf{0},\;
\mathbf{V}_J\succeq \mathbf{0},\;\\
&\eqref{eq:const2}\text{--}\eqref{eq:const7}.
\end{align}
\end{subequations}
Problem~\eqref{eq:mm_surrogate_problem} is still non-convex because $\mathbf{H}_C(\boldsymbol{x}_C)$ and $\mathbf{H}_J(\boldsymbol{x}_J)$ depend on the PA positions. We therefore employ a two-block BCD--SCA procedure.

\emph{Power--radiation update:} Fix $(\boldsymbol{x}_C,\boldsymbol{x}_J)=(\boldsymbol{x}_C^{(k)},\boldsymbol{x}_J^{(k)})$, which renders $\mathbf{H}_C$ and $\mathbf{H}_J$ constant. Then $\hat{\bar R}_C(\cdot\,|\,\boldsymbol{\Psi}^{(k)})$ is concave in $(\mathbf{V}_C,\mathbf{V}_J)$ because it is a sum of logarithms of affine trace expressions plus affine terms. Together with the affine constraints in \eqref{eq:dep_linearized_new} and \eqref{eq:diag_constraints}, the update reduces to a convex SDP, yielding
$(\mathbf{V}_C^{(k+1)},\mathbf{V}_J^{(k+1)},\mathbf{s}_C^{(k+1)},\mathbf{s}_J^{(k+1)},P_C^{(k+1)},P_J^{\max,(k+1)})$.

\emph{PA-position update with proximal SCA:}
Fix the updated power/radiation variables and optimize $(\boldsymbol{x}_C,\boldsymbol{x}_J)$. We linearize the position-dependent terms
$\mathrm{Tr}\big(\mathbf{H}_C\mathbf{V}_C\big)$ and
$\mathrm{Tr}\big(\mathbf{H}_J\mathbf{V}_J\big)$ around
$(\boldsymbol{x}_C^{(k)},\boldsymbol{x}_J^{(k)})$ and add a small proximal regularizer. The resulting position subproblem becomes a convex QP/SOCP \cite{alizadeh2003second} under \eqref{eq:dep_linearized_new} and \eqref{eq:const5}--\eqref{eq:const7}, producing $(\boldsymbol{x}_C^{(k+1)},\boldsymbol{x}_J^{(k+1)})$.

The MM property ensures that $\bar R_C(\boldsymbol{\Psi}^{(k)})$ is non-decreasing across iterations, while the covertness requirement is enforced  (through its affine surrogate) at every iteration. If the SDR yields higher-rank solutions, a rank-one feasible beamformer can be recovered using Gaussian randomization \cite{ma2021robust}.

\subsubsection{Algorithm Summary}

Algorithm~\ref{alg:mm_bcd_sca} summarizes the proposed solver. At each iteration, we update the DEP surrogate (Danskin--SCA), refresh the MM lower bound for the average-rate objective, and then perform the two-block BCD--SCA updates.

\begin{algorithm}[t]
\caption{MM-BCD-SCA Algorithm for Average Covert Rate Maximization}
\label{alg:mm_bcd_sca}
\SetAlgoLined
\DontPrintSemicolon
\KwIn{Initial feasible point $\boldsymbol{\Psi}^{(0)}$; weights $\{\xi_\ell,w_\ell\}_{\ell=1}^L$;
covertness level $\epsilon$; tolerances $\delta_{\mathrm{out}},\delta_{\mathrm{in}}$;
maximum iterations $K_{\max},T_{\max}$.}
\KwOut{Solution $\boldsymbol{\Psi}^{\star}$.}

\While{$k<K_{\max}$}{
Obtain $(\tau^{(k)},g^{(k)},\mathbf{d}_k)$ by using Alg.~\ref{alg:tau_search};\;
Form the affine covertness constraint using \eqref{eq:dep_linearized_new} with $(g^{(k)},\mathbf{d}_k)$;\;

Compute $I(\boldsymbol{\Psi}^{(k)})$ and $D_\ell^{(k)}=\xi_\ell I(\boldsymbol{\Psi}^{(k)})+\sigma_b^2,\ \forall \ell$ as in \eqref{eq:mm_lowerbound};\;

$\boldsymbol{\Psi}^{(k,0)}\leftarrow \boldsymbol{\Psi}^{(k)}$;\;

\While{$t<T_{\max}$}{

Fix $(\boldsymbol{x}_C,\boldsymbol{x}_J)=\left(\boldsymbol{x}_C^{(k,t)},\boldsymbol{x}_J^{(k,t)}\right)$ and
solve the SDP subproblem \eqref{eq:mm_surrogate_problem} with fixed positions to obtain  $\boldsymbol{\Psi}^{(k,t+1)}$;\;

\If{$\mathrm{rank}(\mathbf{V}_C^{(k,t+1)})>1$ \textbf{or} $\mathrm{rank}(\mathbf{V}_J^{(k,t+1)})>1$}{
Extract feasible rank-one beamformers via Gaussian randomization;\;
}

Fix the updated power/radiation variables and solve the convex position subproblem \eqref{eq:mm_surrogate_problem} to update $(\boldsymbol{x}_C^{(k,t+1)},\boldsymbol{x}_J^{(k,t+1)})$;\;

Form $\boldsymbol{\Psi}^{(k,t+1)}$;\;

\If{$\big\|\boldsymbol{\Psi}^{(k,t+1)}-\boldsymbol{\Psi}^{(k,t)}\big\|_2\le \delta_{\mathrm{in}}$}{
\textbf{break};}
$t\leftarrow t+1$;\;
}

$\boldsymbol{\Psi}^{(k+1)}\leftarrow \boldsymbol{\Psi}^{(k,t+1)}$; $R^{(k+1)}\leftarrow \bar R_C(\boldsymbol{\Psi}^{(k+1)})$;\;

\If{$\big|R^{(k+1)}-R^{(k)}\big|\le \delta_{\mathrm{out}}$}{
\textbf{break};}
$k\leftarrow k+1$;\;
}
\Return{$\boldsymbol{\Psi}^{\star}\leftarrow \boldsymbol{\Psi}^{(k+1)}$}\;
\end{algorithm}

\begin{algorithm}[t]
\caption{Optimal Detection Threshold Search Algorithm for $g(\boldsymbol{\Psi})=\min_{\tau>0}P_{\mathrm{dep}}(\tau,\boldsymbol{\Psi})$}
\label{alg:tau_search}
\SetAlgoLined
\DontPrintSemicolon

\KwIn{Current design $\boldsymbol{\Psi}$; piecewise DEP expressions in Section~V.}
\KwOut{$\tau^\star$, $g(\boldsymbol{\Psi})$, and $\mathbf{d}=\nabla_{\boldsymbol{\Psi}}P_{\mathrm{dep}}(\tau^\star,\boldsymbol{\Psi})$.}

Construct the candidate set $\mathcal{T}(\boldsymbol{\Psi})$ by collecting all breakpoints
where local error probabilities switch among $\{0,\text{linear},1\}$ regimes;\;

$\tau^\star \leftarrow \arg\min_{\tau \in \mathcal{T}(\boldsymbol{\Psi})} P_{\mathrm{dep}}(\tau,\boldsymbol{\Psi})$;\;
$g(\boldsymbol{\Psi}) \leftarrow P_{\mathrm{dep}}(\tau^\star,\boldsymbol{\Psi})$;\;

Compute $\mathbf{d}\leftarrow \nabla_{\boldsymbol{\Psi}} P_{\mathrm{dep}}(\tau,\boldsymbol{\Psi})\big|_{\tau=\tau^\star}$\;
\If{multiple minimizers exist}{
Choose $\mathbf{d}$ as any convex combination of the gradients of active minimizers;\;
}

\Return{$(\tau^\star,g(\boldsymbol{\Psi}),\mathbf{d})$}\;
\end{algorithm}

\section{Numerical Results}
\label{sec:numer_results}

In this section, we validate the analytical results and examine how key system parameters affect performance under the three power-radiation models. We also include benchmark schemes to demonstrate the effectiveness of the proposed algorithm. In the simulations, we consider multiple wardens and set the Gaussian noise power at all nodes to $-114$~dBm. Unless otherwise stated, the simulation parameters follow those in the paper and are summarized in Table~\ref{tab:sac_hyperparams}.

\begin{table}[t]
\caption{System Parameter Settings}
\label{tab:sac_hyperparams}
\centering
\renewcommand{\arraystretch}{1.14}
\begin{tabular}{l c l c}
\hline
\textbf{Parameter\&Notion} & \textbf{Value}  \\
\hline
Effective refractive index  $n_{eff}$     & 1.4              \\
Carrier frequency  $f_c$             &  5 GHz           \\
Maximum transmit and jamming power ($P_{max}$,$P^{max}_J)$&  (100, 40)mW     \\
Lateral offset $D$           & 0.4m            \\
Height and length of the waveguide $(L,H)$         & (4m, 4m)                 \\
Bob's position          &[2.1,-0.3,0,0]        \\
Numbers of the PAs on waveguide C and J $(N_1, N_2)$           & (4, 4) \\
\hline
\end{tabular}
\end{table}

\subsection{System DEP vs. Detection Threshold}

\begin{figure}[t]
\centering
\includegraphics[width=0.8\linewidth]{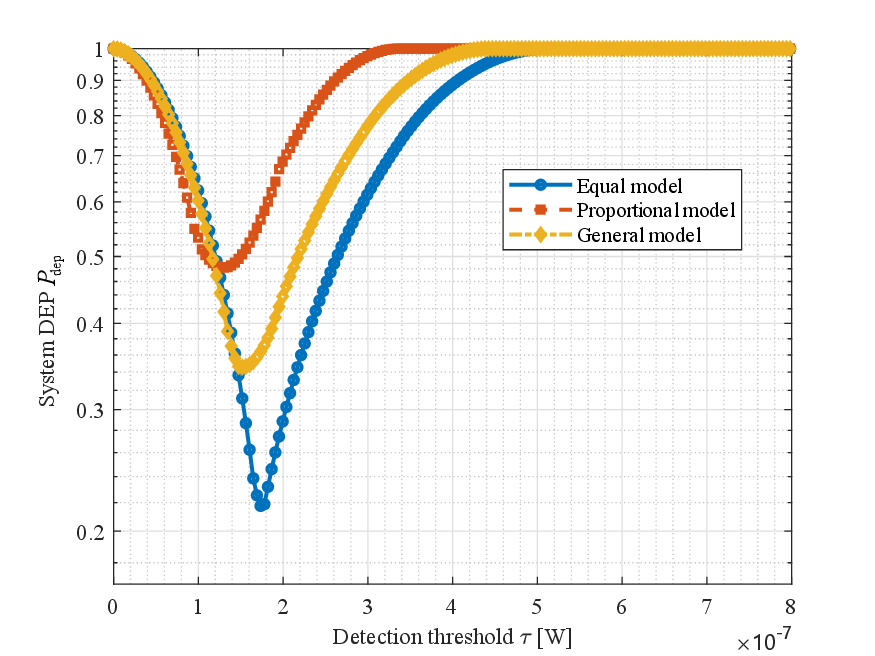}
\caption{System DEP $P_{dep}$ under the variation of detection threshold $\tau$ with $5$ wardens.}
\label{fig:Dep_tau_m=5}
\end{figure}

We first plot Fig.~\ref{fig:Dep_tau_m=5} to study how the warden detection threshold affects the system DEP under the three power–radiation models with $5$ wardens. As expected, $P_{dep}$ exhibits a pronounced U-shaped behavior: when $\tau$ is too small, false alarms dominate and $P_{\mathrm{dep}}$ stays close to one; when $\tau$ is too large, miss detections dominate, again pushing $P_{\mathrm{dep}}$ toward one. Therefore, the optimal operating point occurs at an intermediate threshold that balances these two error events. Moreover, both the location and the depth of the minimum differ across models, indicating that the power-radiation law of the PASS architecture fundamentally alters the wardens' statistical distinguishability. In particular, the equal model yields the deepest minimum in this setting, suggesting that more uniform radiation across PAs can make aggregate energy observations less informative to threshold-based wardens and thus improve system-level covertness under majority voting.

\begin{figure}[t]
\centering
\includegraphics[width=0.8\linewidth]{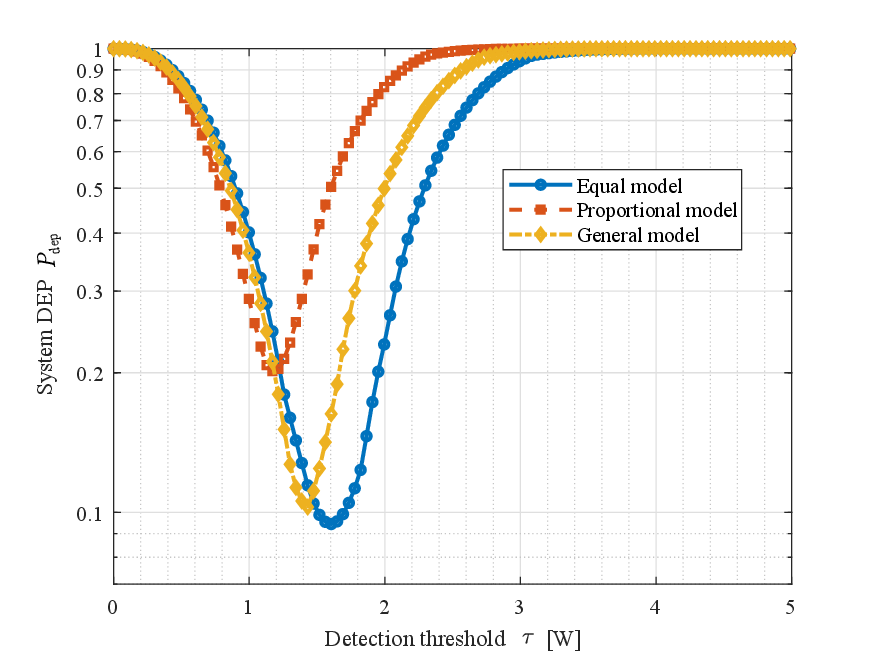}
\caption{System DEP $P_{dep}$ under the variation of detection threshold $\tau$ with $8$ wardens.}
\label{fig:Dep_tau_m=8}
\end{figure}

Fig.~\ref{fig:Dep_tau_m=8} further investigates the impact of scaling the number of wardens. Compared with Fig.~\ref{fig:Dep_tau_m=5}, the curves remain U-shaped but the minima shift and $P_{\mathrm{dep}}$ becomes more sensitive to $\tau$, reflecting the stronger collective decision capability enabled by additional wardens under majority fusion. This trend confirms that system covertness is not solely determined by an individual warden’s local statistic; rather, it is critically governed by the interaction between local detection behavior and the fusion rule. Importantly, the relative performance ordering among the three power–radiation models remains unchanged, suggesting that the proposed PASS-enabled structured radiation/collection mechanism can provide robust covertness benefits even as the adversarial sensing network scales. These observations motivate the subsequent worst-case covertness-constrained design, where the detection threshold is treated adversarially and the optimization explicitly enforces a system-level DEP requirement across $\tau$.

\subsection{System DEP vs. Maximum Jamming Power}

\begin{figure}[t]
\centering
\includegraphics[width=0.8\linewidth]{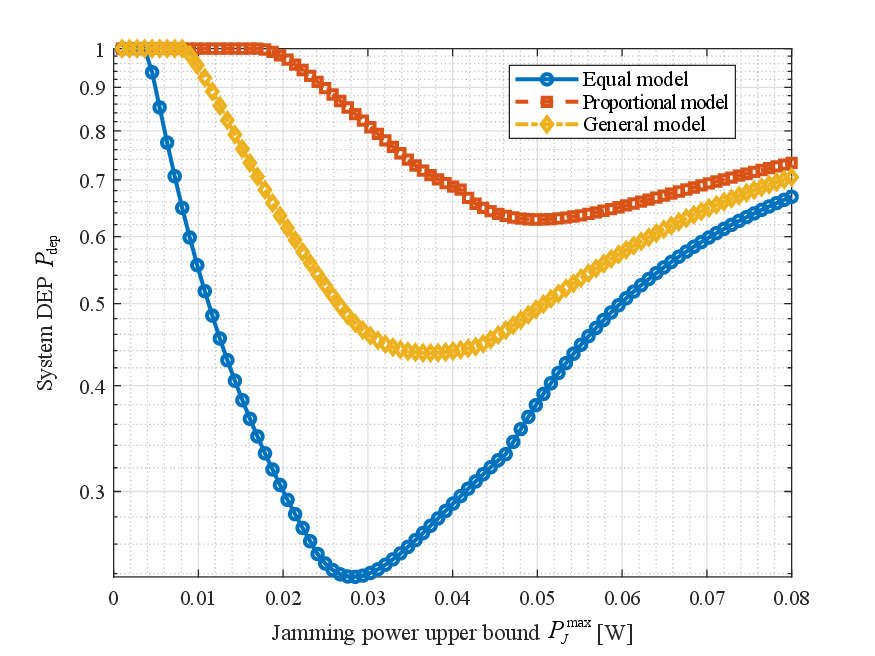}
\caption{System DEP $P_{dep}$ under the variation of maximum jamming power $P^{max}_{J}$ with $5$ wardens.}
\label{fig:Dep_Pj_m=5}
\end{figure}

We plots the system DEP $P_{\mathrm{dep}}$ versus the maximum jamming power budget $P_J^{\max}$ under the three power--radiation models in Fig.~\ref{fig:Dep_Pj_m=5}. A clear non-monotonic behavior is observed: when $P_J^{\max}$ is small, the jammer cannot sufficiently mask the covert transmission and $P_{\mathrm{dep}}$ remains high; as $P_J^{\max}$ increases, $P_{\mathrm{dep}}$ decreases and reaches a minimum, indicating that moderate jamming effectively blurs the wardens' energy statistics. However, further increasing $P_J^{\max}$ can degrade covertness and lead to a higher $P_{\mathrm{dep}}$. This result highlights that \emph{more jamming is not always better} in PASS-enabled systems, because excessive radiated jamming power may itself become a strong detection cue under structured PA deployment. Moreover, the equal model consistently achieves the lowest $P_{\mathrm{dep}}$ over a wide range of $P_J^{\max}$, suggesting that uniform radiation across PAs mitigates the detectability of the composite signal-jamming footprint under majority fusion.

\begin{figure}[t]
\centering
\includegraphics[width=0.8\linewidth]{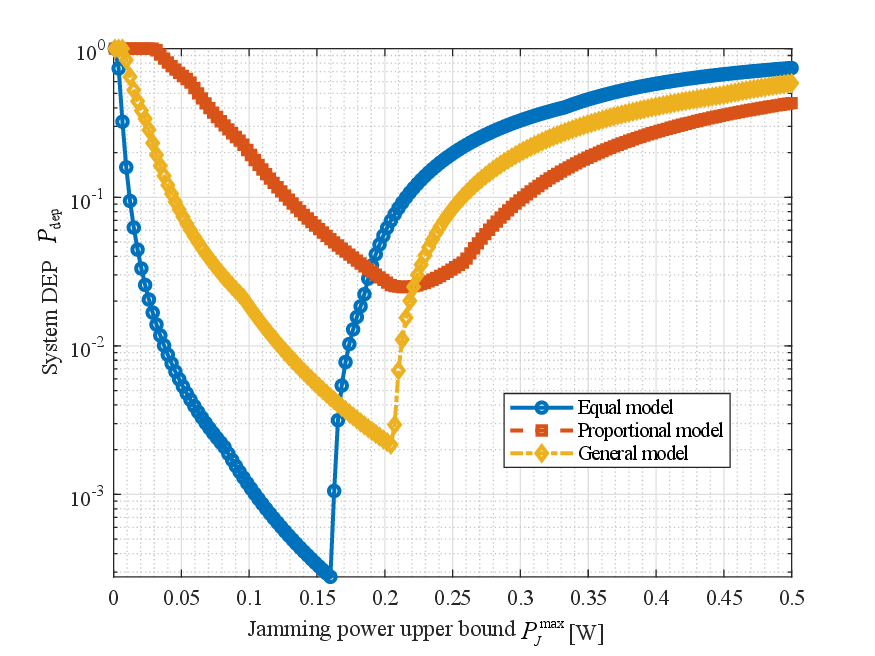}
\caption{System DEP $P_{dep}$ under the variation of maximum jamming power $P^{max}_{J}$ with $8$ wardens.}
\label{fig:Dep_Pj_m=8}
\end{figure}

Fig.~\ref{fig:Dep_Pj_m=8}  studies the same relationship in a denser adversarial setting, where the $y$-axis is plotted in logarithmic scale to better reveal the error-floor behavior. The curves preserve the non-monotonic structure, while the minima become sharper and shift in $P_J^{\max}$, implying that more wardens strengthen collective detection and make the system more sensitive to jamming-power mismatch. Notably, the minimum $P_{\mathrm{dep}}$ can drop by orders of magnitude in the moderate-$P_J^{\max}$ regime, indicating that the jamming budget should be carefully calibrated as the sensing network scales. Meanwhile, the relative ordering among the three power-radiation models remains consistent, corroborating that the proposed PASS architecture and  power-radiation modeling provide robust guidance for system-level covertness design. These results further motivate a worst-case covertness-constrained optimization that jointly tunes the structured PA radiation and jamming strategy, rather than relying on monotone power-increase heuristics.

\subsection{Average Covert Rate vs. Covert Transmit Power}

\begin{figure}[t]
\centering
\includegraphics[width=0.8\linewidth]{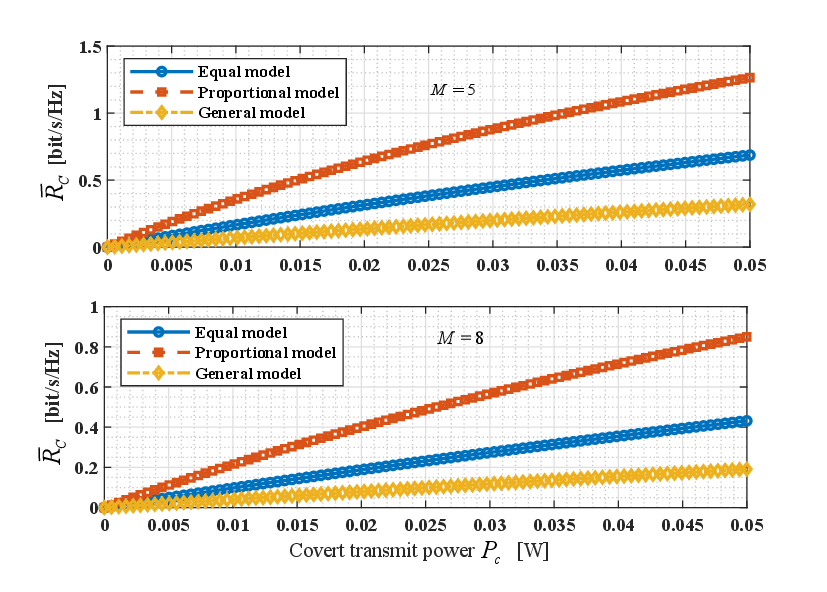}
\caption{Average covert rate $\bar{R}_c$ under the variation of covert transmit power $P_{c}$.}
\label{fig:ACR}
\end{figure}

Fig.~\ref{fig:ACR} illustrates the average covert rate $\bar{R}_c$ as a function of the covert transmit power $P_c$ for different number wardens. As expected, $\bar{R}_c$ increases monotonically with $P_c$ in both cases, since higher transmit power improves Bob's received SNR and thus increases the achievable rate. Comparing the two subplots, increasing the number of wardens from $M=5$ to $M=8$ consistently reduces $\bar{R}_c$ for a given $P_c$, reflecting the tighter system-level covertness requirement imposed by a larger adversarial sensing network. Moreover, the three power-radiation models yield clearly distinct rate slopes: the proportional model achieves the highest $\bar{R}_c$ over the entire $P_c$ range, followed by the equal model, while the general model results in the smallest $\bar{R}_c$. This indicates that enforcing structured PA radiation via deterministic power fractions can significantly shape the effective channel gain at Bob and, consequently, the covert throughput. Overall, Fig.~\ref{fig:ACR} corroborates the necessity of explicitly accounting for the underlying power-radiation law when optimizing $\boldsymbol{\Psi}$, as different radiation models can lead to substantially different covert-rate gains under the same transmit-power budget and warden configuration.

\subsection{Maximum Average covert Rate}

\begin{figure}[t]
\centering
\includegraphics[width=1\linewidth]{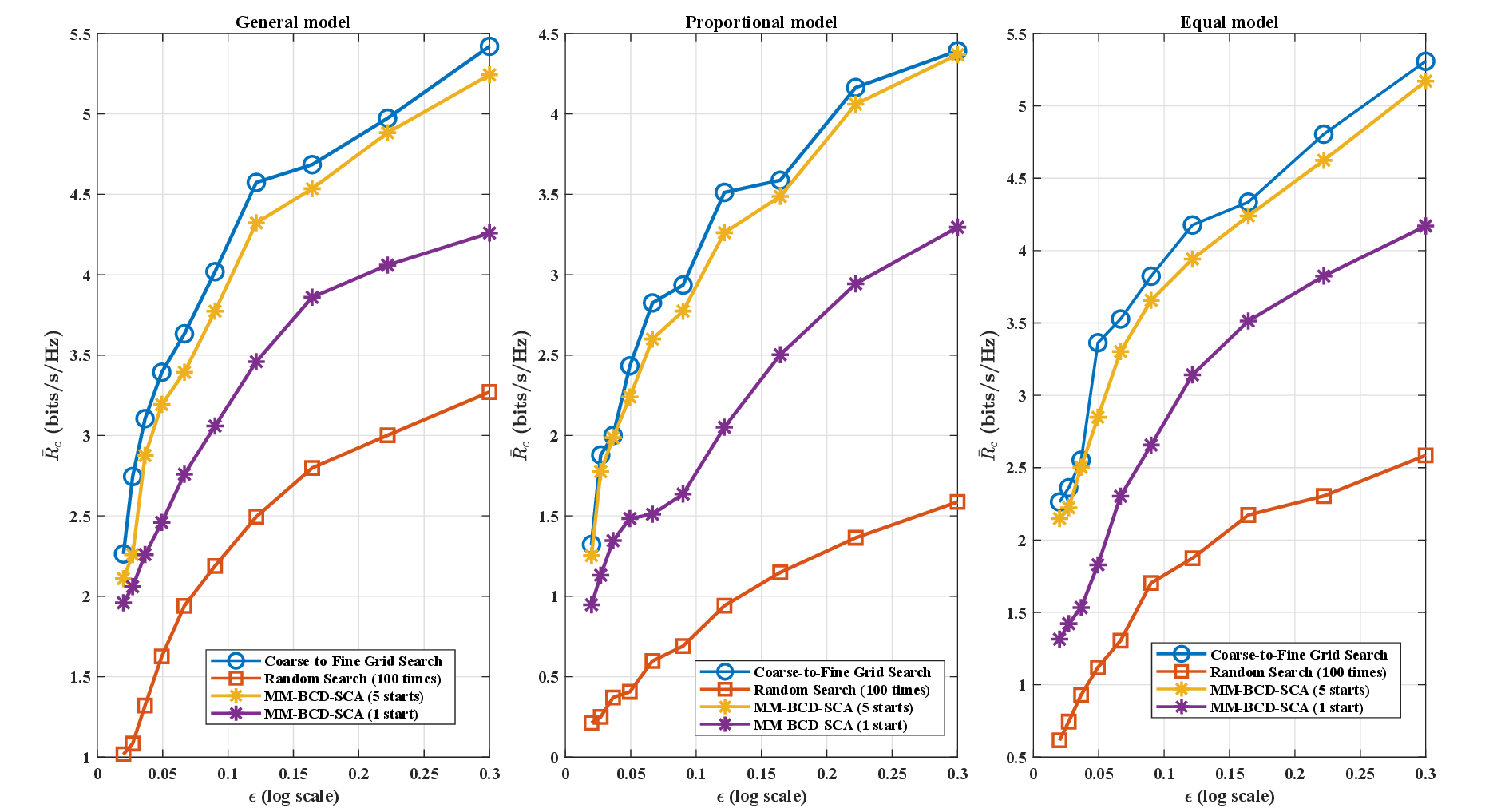}
\caption{The comparison results of maximum $\bar{R}_c$ under the variation of the covertness constraints with $5$ wardens.}
\label{fig:max_acr_m=5}
\end{figure}

\begin{figure}[t]
\centering
\includegraphics[width=1\linewidth]{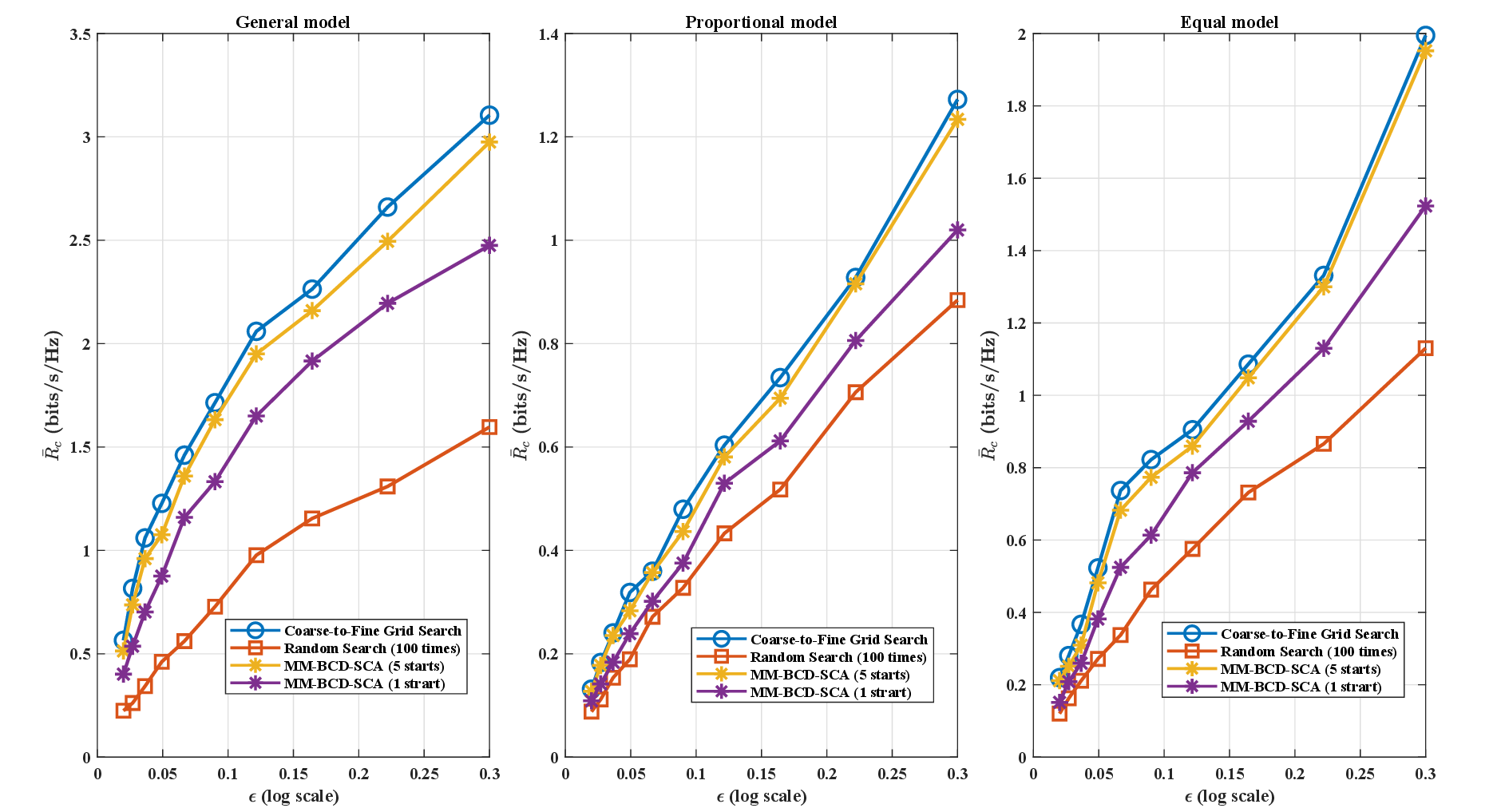}
\caption{The comparison results of maximum $\bar{R}_c$ under the variation of the covertness constraints with $8$ wardens.}
\label{fig:max_acr_m=8}
\end{figure}

To evaluate the proposed MM–BCD–SCA method, we consider two baselines: a coarse-to-fine grid search and a random search. Since the design variables $\boldsymbol{\Psi}$ is high-dimensional and continuous, a full-dimensional exhaustive search is computationally prohibitive. We thus adopt a coarse-to-fine benchmark that performs a coarse grid over the PA placement variables $(\boldsymbol{x}_C, \boldsymbol{x}_J)$and, for each placement candidate, conducts a fine power grid scan over $(P_C, P_J^{\max})$ Satisfying the power constraint \eqref{eq:const2}. This benchmark provides a strong yet tractable reference within the discretized placement space. We also include a random baseline: for each $\epsilon$, we independently generate $100$ feasible designs, compute $\bar{R}_c$ for each trial, and report the average $\bar{R}_c$ over all trials. This baseline reflects the performance of uninformed sampling without optimization and provides a conservative reference.

From Figs.~\ref{fig:max_acr_m=5} and~\ref{fig:max_acr_m=8}, we can observe that the proposed method consistently outperforms random search by a substantial margin, demonstrating that uninformed feasible sampling cannot effectively exploit the coupled placement–power–beamforming design space. Moreover, increasing the number of multi-start initializations from $K=1$ to $K=5$ yields noticeable gains and brings the proposed solution close to the coarse-to-fine grid-search benchmark, indicating that the alternating MM–BCD–SCA framework can efficiently approach high-quality stationary points with limited additional overhead. Comparing Figs.~\ref{fig:max_acr_m=5} and~\ref{fig:max_acr_m=8}, increasing the number of wardens from $M=5$ to $M=8$ consistently reduces $\bar{R}_c$ across all $\epsilon$ and all radiation models. This behavior is expected because more wardens strengthen the overall detection capability under majority voting, thereby tightening the effective covertness constraint and forcing more conservative power/radiation decisions. Nevertheless, the proposed MM–BCD–SCA method remains robust as $M$ increases and continues to closely track the grid-search reference, demonstrating its effectiveness and scalability with respect to the warden density.


\section{CONCLUSIONS}
\label{sec:conclusion}

In this paper, we have explored the integration of PASS technology with covert communication to counter distributed surveillance. By utilizing a dual-waveguide architecture, we demonstrated how the spatial programmability of PAs and structured radiation laws can be exploited to mask transmission from multiple cooperative wardens. Our theoretical analysis successfully addressed the challenge of non-identically distributed warden statistics under a majority-voting fusion rule, resulting in a comprehensive piecewise characterization of the system-level DEP. Furthermore, the proposed MM-BCD-SCA algorithm effectively handles the bilinear coupling between baseband power and physical antenna placement, converging to high-quality solutions that closely approach global grid-search benchmarks. Our findings indicate that the system-level DEP is highly sensitive to the interaction between the radiation model and the fusion rule, with the equal radiation model providing superior covertness in dense warden scenarios. These results underscore the potential of PASS as a geometry-driven security paradigm for 6G networks, offering robust protection in complex, interference-limited environments.

\bibliographystyle{IEEEtran}
\bibliography{IEEE_trans}

\end{document}